\documentclass[12pt,english]{article}
\usepackage{times}
\usepackage[T1]{fontenc}
\usepackage{geometry}
\usepackage{array}
\usepackage{rotating}
\usepackage{color}
\usepackage{graphicx}
\usepackage{setspace}
\usepackage{amssymb}

\makeatletter

\providecommand{\tabularnewline}{\\}

\usepackage{bm}

\usepackage{babel}
\makeatother
\begin{document}

\section*{\noindent Inverse Source Method for Constrained Phased Array Synthesis
through Null-Space Exploitation}

\noindent ~

\noindent \vfill

\noindent L. Poli,$^{(1)(2)}$ \emph{Senior Member, IEEE}, P. Rocca,$^{(1)(2)(3)}$
\emph{Fellow, IEEE}, \emph{}A. Benoni,$^{(1)(2)}$ \emph{Member},
\emph{IEEE}, and A. Massa,$^{(1)(2)(4)(5)(6)}$ \emph{Fellow, IEEE}

\noindent \vfill

\noindent {\footnotesize ~}{\footnotesize \par}

\noindent {\scriptsize $^{(1)}$} \emph{\scriptsize ELEDIA Research
Center} {\scriptsize (}\emph{\scriptsize ELEDIA}{\scriptsize @}\emph{\scriptsize UniTN}
{\scriptsize - University of Trento)}{\scriptsize \par}

\noindent {\scriptsize DICAM - Department of Civil, Environmental,
and Mechanical Engineering}{\scriptsize \par}

\noindent {\scriptsize Via Mesiano 77, 38123 Trento - Italy}{\scriptsize \par}

\noindent \textit{\emph{\scriptsize E-mail:}} {\scriptsize \{}\emph{\scriptsize lorenzo.poli,
paolo.rocca, arianna.benoni, andrea.massa}{\scriptsize \}@}\emph{\scriptsize unitn.it}{\scriptsize \par}

\noindent {\scriptsize Website:} \emph{\scriptsize www.eledia.org/eledia-unitn}{\scriptsize \par}

\noindent {\scriptsize ~}{\scriptsize \par}

\noindent {\scriptsize $^{(2)}$} \emph{\scriptsize CNIT - \char`\"{}University
of Trento\char`\"{} ELEDIA Research Unit }{\scriptsize \par}

\noindent {\scriptsize Via Mesiano 77, 38123 Trento - Italy}{\scriptsize \par}

\noindent {\scriptsize Website:} \emph{\scriptsize www.eledia.org/eledia-unitn}{\scriptsize \par}

\noindent {\scriptsize ~}{\scriptsize \par}

\noindent {\scriptsize $^{(3)}$} \emph{\scriptsize ELEDIA Research
Center} {\scriptsize (}\emph{\scriptsize ELEDIA}{\scriptsize @}\emph{\scriptsize XIDIAN}
{\scriptsize - Xidian University)}{\scriptsize \par}

\noindent {\scriptsize P.O. Box 191, No.2 South Tabai Road, 710071
Xi'an, Shaanxi Province - China }{\scriptsize \par}

\noindent {\scriptsize E-mail:} \emph{\scriptsize paolo.rocca@xidian.edu.cn}{\scriptsize \par}

\noindent {\scriptsize Website:} \emph{\scriptsize www.eledia.org/eledia-xidian}{\scriptsize \par}

\noindent {\scriptsize ~}{\scriptsize \par}

\noindent {\scriptsize $^{(4)}$} \emph{\scriptsize ELEDIA Research
Center} {\scriptsize (}\emph{\scriptsize ELEDIA}{\scriptsize @}\emph{\scriptsize UESTC}
{\scriptsize - UESTC)}{\scriptsize \par}

\noindent {\scriptsize School of Electronic Science and Engineering,
Chengdu 611731 - China}{\scriptsize \par}

\noindent \textit{\emph{\scriptsize E-mail:}} \emph{\scriptsize andrea.massa@uestc.edu.cn}{\scriptsize \par}

\noindent {\scriptsize Website:} \emph{\scriptsize www.eledia.org/eledia}{\scriptsize -}\emph{\scriptsize uestc}{\scriptsize \par}

\noindent {\scriptsize ~}{\scriptsize \par}

\noindent {\scriptsize $^{(5)}$} \emph{\scriptsize ELEDIA Research
Center} {\scriptsize (}\emph{\scriptsize ELEDIA@TSINGHUA} {\scriptsize -
Tsinghua University)}{\scriptsize \par}

\noindent {\scriptsize 30 Shuangqing Rd, 100084 Haidian, Beijing -
China}{\scriptsize \par}

\noindent {\scriptsize E-mail:} \emph{\scriptsize andrea.massa@tsinghua.edu.cn}{\scriptsize \par}

\noindent {\scriptsize Website:} \emph{\scriptsize www.eledia.org/eledia-tsinghua}{\scriptsize \par}

\noindent {\footnotesize ~}{\footnotesize \par}

\noindent {\footnotesize $^{(6)}$} {\scriptsize School of Electrical
Engineering}{\scriptsize \par}

\noindent {\scriptsize Tel Aviv University, Tel Aviv 69978 - Israel}{\scriptsize \par}

\noindent \textit{\emph{\scriptsize E-mail:}} \emph{\scriptsize andrea.massa@eng.tau.ac.il}{\scriptsize \par}

\noindent {\scriptsize Website:} \emph{\scriptsize https://engineering.tau.ac.il/}{\scriptsize \par}

\noindent \emph{This work has been submitted to the IEEE for possible
publication. Copyright may be transferred without notice, after which
this version may no longer be accessible.}

\newpage
\section*{Inverse Source Method for Constrained Phased Array Synthesis through
Null-Space Exploitation}

\textcolor{black}{~}

\noindent \textcolor{black}{~}

\noindent \textcolor{black}{~}

\begin{flushleft}L. Poli, P. Rocca, A. Benoni, and A. Massa\end{flushleft}

\textcolor{black}{\vfill}

\begin{abstract}
\noindent A versatile approach for the synthesis of phased array (\emph{PA})
\textcolor{black}{antennas able to fit user-defined power pattern
masks, while fulfilling additional geometrical and/or electrical constraints}
on the geometry of the array aperture \textcolor{black}{and/or on
the array excitations is presented. Such a synthesis method is based
on the} \textcolor{black}{\emph{inverse source}} \textcolor{black}{(}\textcolor{black}{\emph{IS}}\textcolor{black}{)
formulation and exploits the null-space of the radiation operator
that causes the non-uniqueness of the} \textcolor{black}{\emph{IS}}
\textcolor{black}{problem at hand. More in detail, the unknown element
excitations of the} \textcolor{black}{\emph{PA}} \textcolor{black}{are
expressed as the linear combination of a} \textcolor{black}{\emph{minimum-norm}}
\textcolor{black}{or} \textcolor{black}{\emph{radiating}} \textcolor{black}{(}\textcolor{black}{\emph{RA}}\textcolor{black}{)
term and a suitable} \textcolor{black}{\emph{non-radiating}} \textcolor{black}{(}\textcolor{black}{\emph{NR}}\textcolor{black}{)
component. The former, computed via the truncated singular value decomposition
(}\textcolor{black}{\emph{SVD}}\textcolor{black}{) of the array radiation
operator, is devoted to generate a far-field power pattern that fulfills
user-defined pattern masks. The other one belongs to the null-space
of the radiation operator and allows one to fit additional geometrical
and/or electrical constraints on the geometry of the array aperture
and/or on the beam-forming network (}\textcolor{black}{\emph{BFN}}\textcolor{black}{)
when determined with a customized global optimization strategy. A
set of numerical examples, concerned with various array arrangements
and additional design targets, is reported to prove the effectiveness
of the proposed approach.}

\noindent \textcolor{black}{\vfill}
\end{abstract}
\noindent \textbf{\textcolor{black}{Key words}}\textcolor{black}{:
Phased Array Antenna (}\textcolor{black}{\emph{PA}}\textcolor{black}{),
Power Pattern Synthesis, Inverse Problem (}\textcolor{black}{\emph{IP}}\textcolor{black}{),
Inverse Source (}\textcolor{black}{\emph{IS}}\textcolor{black}{),
Singular Value Decomposition (}\textcolor{black}{\emph{SVD}}\textcolor{black}{),
Optimization Method.}

\newpage
\section{\textcolor{black}{Introduction}}

\noindent \textcolor{black}{Phased array (}\textcolor{black}{\emph{PA}}\textcolor{black}{)
antennas have become pivotal devices for high-performance electromagnetic
(}\textcolor{black}{\emph{EM}}\textcolor{black}{) systems across a
wide spectrum of applications including radar, satellite communications,
wireless networks, remote sensing, and navigation \cite{Haupt 2015}.
The capability to produce electronically steerable beams with high
precision and gain, combined with rapid reconfigurability, makes}
\textcolor{black}{\emph{PA}}\textcolor{black}{s fundamental assets
in both traditional and emerging scenarios where adaptability, efficiency,
and compactness are key-issues. As the global demand for connected,
intelligent, and resilient infrastructures increases,} \textcolor{black}{\emph{PA}}\textcolor{black}{s
have a central role in non-conventional domains such as autonomous
mobility, smart buildings, and distributed sensing platforms. Within
this framework, the design of innovative} \textcolor{black}{\emph{PA}}
\textcolor{black}{solutions with tailored geometrical layouts and
simplified feeding networks is of paramount interest \cite{Herd 2016}\cite{Rocca 2016}.}

\noindent \textcolor{black}{The synthesis of} \textcolor{black}{\emph{PA}}
\textcolor{black}{antennas - specifically, the determination of suitable
excitations to generate prescribed far-field radiation patterns -
is still a core challenge. Traditionally,} \textcolor{black}{\emph{PA}}\textcolor{black}{s
synthesis techniques have been mainly focused on fulfilling radiation
requirements (namely, {}``}\textcolor{black}{\emph{functional}}\textcolor{black}{''
requirements) such as matching a target power pattern or minimizing
the sidelobe level (}\textcolor{black}{\emph{SLL}}\textcolor{black}{)
\cite{Elliott 2003}-\cite{Mailloux 2018} or fitting a power pattern
mask \cite{Haupt 2010}\cite{Mailloux 2018}. However, to address
the needs of practical implementations, it is increasingly important
to incorporate in the design process additional geometrical and electrical
constraints also indicated as {}``}\textcolor{black}{\emph{non-functional}}\textcolor{black}{''
requirements. They may include enforcing modular or sparse array topologies
\cite{Anselmi 2023}\cite{Chen 2023}, excluding feed points in a
set of regions of the} \textcolor{black}{\emph{PA}} \textcolor{black}{aperture
or according to thinned configurations \cite{Lee 2024}\cite{Poli 2025},
or ensuring the compatibility with low-cost and low-power beam-forming
networks (}\textcolor{black}{\emph{BFN}}\textcolor{black}{s) implementations
\cite{Rocca 2019}\cite{Zhu 2024}. Unfortunately, most conventional
approaches to} \textcolor{black}{\emph{PA}} \textcolor{black}{synthesis
neglect or do not inherently deal with these constraints.}

\noindent \textcolor{black}{To overcome these limitations, this work
proposes a novel approach for the synthesis of} \textcolor{black}{\emph{PA}}\textcolor{black}{s
based on an inverse source (}\textcolor{black}{\emph{IS}}\textcolor{black}{)
formulation \cite{Bertero 1998} where the desired radiation pattern
and the element excitations are the data and the unknowns of the problem
at hand, respectively. By modeling the data-unknowns relationship
with the linear radiation operator, according to the Green's function
theory \cite{Kong 2008}, the} \textcolor{black}{\emph{PA}} \textcolor{black}{synthesis
is cast as the inverse problem (}\textcolor{black}{\emph{IP}}\textcolor{black}{)
of determining an excitation vector that fits far-field power constraints.
It is well-known that} \textcolor{black}{\emph{IS}} \textcolor{black}{problems
are ill-posed \cite{Chew 1994}-\cite{Salucci 2018a} owing to the
non-uniqueness of the solution, which is a direct consequence of the
presence of non-radiating (}\textcolor{black}{\emph{NR}}\textcolor{black}{)
current components, namely} \textcolor{black}{\emph{PA}} \textcolor{black}{excitations
that radiate only within the antenna support and do not contribute
to the far-field pattern \cite{Devaney 1973}-\cite{Marengo 2000}.}

\noindent \textcolor{black}{Unlike inverse scattering problems, the
non-uniqueness feature of this} \textcolor{black}{\emph{IS}} \textcolor{black}{problem
is not a limitation/issue, but an additional degree of freedom (}\textcolor{black}{\emph{DoF}}\textcolor{black}{)
\cite{Bucci 1989} to fulfill} \textcolor{black}{\emph{non-functional}}
\textcolor{black}{design constraints. More in detail, the} \textcolor{black}{\emph{PA}}
\textcolor{black}{element excitations are expressed as the superposition
of two components, namely a minimum-norm or radiating (}\textcolor{black}{\emph{RA}}\textcolor{black}{)
term, which fulfills the} \textcolor{black}{\emph{functional}} \textcolor{black}{radiation
objective, and a tailored} \textcolor{black}{\emph{NR}} \textcolor{black}{component
that allows one to fit additional geometrical and/or electrical constraints.
These latter may include, for example, either the presence of {}``}\textcolor{black}{\emph{forbidden}}\textcolor{black}{''
regions in the} \textcolor{black}{\emph{PA}} \textcolor{black}{aperture
where the excitations are forced to be null or promoting the} \textcolor{black}{\emph{BFN}}
\textcolor{black}{sparsity. The} \textcolor{black}{\emph{NR}} \textcolor{black}{component
is determined through a proper customization of a constrained optimization
approach successfully adopted for the design of both surface currents
of reflectarrays \cite{Salucci 2018b}\cite{Salucci 2022} and inexpensive
static passive} \textcolor{black}{\emph{EM}} \textcolor{black}{skins
(}\textcolor{black}{\emph{SP-EMS}}\textcolor{black}{) \cite{Oliveri 2025}
in the framework of Smart Electromagnetic Environments for next generation
communication systems \cite{Massa 2021}\cite{Yang 2022}. The result
is the definition of a flexible and computationally-efficient synthesis
method for the delivery of physically meaningful and implementable}
\textcolor{black}{\emph{PA}} \textcolor{black}{layouts without compromising
radiation performance. It is worthwhile to point out that the analytical
formulation of the} \textcolor{black}{\emph{IS}} \textcolor{black}{problem
at hand enables the definition of suitable representation bases for
the} \textcolor{black}{\emph{NR}} \textcolor{black}{components \cite{Habashy 1994}\cite{Rocca 2009b}
as well as an effective use of a multi-agent global optimization strategy
to determine the} \textcolor{black}{\emph{NR}} \textcolor{black}{expansion
coefficients by avoiding local minima traps and reducing the overall
computational complexity \cite{Rocca 2009a}.}

\noindent \textcolor{black}{To the best of the authors' knowledge,
the main novelties of this work over the existing state-of-the-art
literature on} \textcolor{black}{\emph{PA}} \textcolor{black}{synthesis
include} (\emph{i}) a new theoretical framework for the synthesis
of \emph{PA} antennas that exploits the non-uniqueness of the \emph{IS}
problem at hand to explicitly take advantage of the components of
the null-space of the radiation operator, (\emph{ii}) an innovative
method to define the \emph{NR} components of the element excitations
for fulfilling user-defined or application-driven geometrical and/or
electrical constraints, (\emph{iii}) the numerical assessment of the
effectiveness and the versatility of the proposed \emph{NR}-based
array synthesis method when applied to various array arrangements,
including linear and planar arrays, and \textcolor{black}{design targets
such as the presence of {}``}\textcolor{black}{\emph{forbidden}}\textcolor{black}{''
regions within the antenna aperture as well as the simplification
of the} \textcolor{black}{\emph{BFN}}\textcolor{black}{.}

\noindent \textcolor{black}{The outline of the paper is as follows.
The problem is mathematically formulated in Sect. \ref{sec:Mathematical-Formulation},
while the proposed} \textcolor{black}{\emph{NR}} \textcolor{black}{synthesis
method is presented in Sect. \ref{sec:Non-Radiating-Current-Based}.
Section \ref{sec:Numerical-Results} is devoted to the numerical analysis
and the assessment of the proposed approach. Eventually, some conclusions
and final remarks are drawn (Sect. \ref{sec:Conclusions}).}

\section{\noindent \textcolor{black}{Mathematical Formulation \label{sec:Mathematical-Formulation}}}

\noindent \textcolor{black}{Let us consider a planar} \textcolor{black}{\emph{PA}}
\textcolor{black}{with aperture $A$ lying on the $xy$ plane and
composed of $N$ elements centered at \{$\left(x_{n},y_{n}\right)$;
$n=1,...,N$\} as shown in Fig. 1. The far-field pattern of the array
at the carrier frequency $f$ is given by\begin{equation}
\mathbf{E}\left(\theta,\phi\right)=\sum_{n=1}^{N}w_{n}\mathbf{a}_{n}\left(\theta,\phi\right)e^{j\frac{2\pi}{\lambda}\left(x_{n}\sin\theta\cos\phi+y_{n}\sin\theta\sin\phi\right)}\label{eq:_EM.field}\end{equation}
where $w_{n}=\alpha_{n}e^{j\beta_{n}}$ is the complex excitation,
with amplitude $\alpha_{n}$ and phase $\beta_{n}$, of the $n$-th
($n=1,...,N$) radiating element of the} \textcolor{black}{\emph{PA}}\textcolor{black}{,
which is characterized by an active element pattern $\mathbf{a}_{n}\left(\theta,\phi\right)$.
Moreover, $\lambda$ is the wavelength at $f$ and $\left(\theta,\phi\right)$
is the angular direction, $\theta$ and $\phi$ being the angular
coordinates ranging within $\theta\in\left[-90:90\right]$ {[}deg{]}
and $\phi\in\left[0:180\right]$ {[}deg{]}, respectively. In order
to assess the feasibility and performance of the array architecture
\cite{Isernia 1998} without any} \textcolor{black}{\emph{a-priori}}
\textcolor{black}{assumption on the type of radiating elements in
the} \textcolor{black}{\emph{PA}}\textcolor{black}{, let us consider
ideal isotropic radiators in absence of mutual coupling, thus $\mathbf{a}_{n}\left(\theta,\phi\right)=\frac{1}{\sqrt{2}}\left(\widehat{\mathbf{\theta}}+\widehat{\mathbf{\phi}}\right)$
($n=1,...,N$), $\widehat{\mathbf{\theta}}$ and $\widehat{\mathbf{\phi}}$
being unit vectors. Accordingly, the expression in (\ref{eq:_EM.field})
simplifies to the array factor\begin{equation}
AF\left(\theta,\left.\phi\right|\underline{r},\underline{w}\right)=\sum_{n=1}^{N}w_{n}e^{j\frac{2\pi}{\lambda}\left(x_{n}\sin\theta\cos\phi+y_{n}\sin\theta\sin\phi\right)},\label{eq:_array.factor}\end{equation}
which is only function of the electrical and geometrical array parameters,
namely the excitations set, $\underline{w}=\left\{ w_{n};n=1,...,N\right\} $,
and the element positions set, $\underline{r}=\left\{ \mathbf{r}_{n}=\left(x_{n},\, y_{n}\right);n=1,...,N\right\} $.}

\noindent \textcolor{black}{In this framework, the formulation of
the {}``standard'' array synthesis problem can be stated as follows:}

\begin{quote}
\textbf{\textcolor{black}{Standard Array Synthesis (}}\textbf{\textcolor{black}{\emph{SAS}}}\textbf{\textcolor{black}{)
Problem}} \textcolor{black}{- Given the upper, $UM\left(\theta,\phi\right)$,
and the lower, $LM\left(\theta,\phi\right)$, power pattern masks
and the positions of the $N$ array elements, $\underline{r}$, define
the amplitude, $\underline{\alpha}=\left\{ \alpha_{n};n=1,...,N\right\} $,
and the phase, $\underline{\beta}=\left\{ \beta_{n};n=1,...,N\right\} $,
values of the set of complex excitations, $\underline{w}$, of the
array so that the radiated power pattern\begin{equation}
P\left(\theta,\left.\phi\right|\underline{w}\right)=\left|AF\left(\theta,\left.\phi\right|\underline{w}\right)\right|^{2}\label{eq:_power.pattern}\end{equation}
minimizes the following} \textcolor{black}{\emph{mask-matching}} \textcolor{black}{metric\begin{equation}
\begin{array}{c}
\Phi_{M}\left(\theta,\left.\phi\right|\underline{w}\right)=\frac{1}{2\pi}\int_{0}^{2\pi}\int_{0}^{\frac{\pi}{2}}\left\{ \left[P\left(\theta,\left.\phi\right|\underline{w}\right)-UM\left(\theta,\phi\right)\right]\, H\left\{ P\left(\theta,\left.\phi\right|\underline{w}\right)-UM\left(\theta,\phi\right)\right\} \right.+\\
\left[LM\left(\theta,\phi\right)-P\left(\theta,\left.\phi\right|\underline{w}\right)\right]\,\left.H\left\{ LM\left(\theta,\phi\right)-P\left(\theta,\left.\phi\right|\underline{w}\right)\right\} \right\} \sin\theta d\theta d\phi\end{array}\label{eq:_mask.matching}\end{equation}
aimed at quantifying the violation of $P\left(\theta,\left.\phi\right|\underline{w}\right)$
from the upper, $UM\left(\theta,\phi\right)$, and the lower, $LM\left(\theta,\phi\right)$,
pattern masks, $H\left\{ \circ\right\} $ being the Heaviside step
function ($H\left\{ \circ\right\} =1$ if $\circ\geq0$ and $H\left\{ \circ\right\} =0$,
otherwise).}
\end{quote}
\noindent \textcolor{black}{Various synthesis methods have been successfully
developed to address the} \textcolor{black}{\emph{SAS}} \textcolor{black}{problem
\cite{Elliott 2003}-\cite{Mailloux 2018}, but even though the arising
excitation set typically ensures the compliance with the power mask
- provided a physical solution exists - it does not account for additional
constraints.}

\noindent \textcolor{black}{By leveraging on the ill-posedness of
the} \textcolor{black}{\emph{IS}} \textcolor{black}{problems and the
non-uniqueness of their solutions, the element excitations can be
expressed as\begin{equation}
\underline{w}=\underline{w}^{RA}+\underline{w}^{NR}\label{eq:_excitation.MN-plus-NR}\end{equation}
where $\underline{w}^{RA}$ ($\underline{w}^{RA}=\left\{ w_{n}^{RA};n=1,...,N\right\} $)
is the set of the minimum-norm weights aimed at generating the far-field
pattern (\ref{eq:_array.factor}) to fulfill the} \textcolor{black}{\emph{functional}}\textcolor{black}{/primary
problem objective by fitting the mask-matching requirement (\ref{eq:_mask.matching})
{[}i.e., $\Phi_{M}\left(\underline{w}^{RA}\right)=0${]}, while $\underline{w}^{NR}$
($\underline{w}^{NR}=\left\{ w_{n}^{NR};n=1,...,N\right\} $) is the
set of non-radiating coefficients generating a null} \textcolor{black}{\emph{EM}}
\textcolor{black}{field outside the aperture $A$, that is\begin{equation}
0=\sum_{n=1}^{N}w_{n}^{NR}e^{j\frac{2\pi}{\lambda}\left(x_{n}\sin\theta\cos\phi+y_{n}\sin\theta\sin\phi\right)}\,.\label{eq:_NR.field.condition}\end{equation}
Therefore, it turns out that $AF\left(\theta,\left.\phi\right|\underline{w}\right)=AF^{RA}\left(\theta,\phi\right)$,
$AF^{RA}\left(\theta,\phi\right)$ being the array factor (\ref{eq:_array.factor})
when $\underline{w}=\underline{w}^{RA}$ {[}i.e., $AF^{RA}\left(\theta,\phi\right)\triangleq\sum_{n=1}^{N}w_{n}^{RA}e^{j\frac{2\pi}{\lambda}\left(x_{n}\sin\theta\cos\phi+y_{n}\sin\theta\sin\phi\right)}${]}.
Accordingly, the constrained version of the} \textcolor{black}{\emph{SAS}}
\textcolor{black}{problem is formulated as follows:}

\begin{quotation}
\noindent \textbf{\textcolor{black}{Constrained Array Synthesis (}}\textbf{\textcolor{black}{\emph{CAS}}}\textbf{\textcolor{black}{)
Problem}} \textcolor{black}{- Given an array of $N$ elements having
positions $\underline{r}$ and excitations $\underline{w}^{ref}$
that affords a reference power pattern $P^{ref}\left(\theta,\phi\right)$
{[}$P^{ref}\left(\theta,\phi\right)\triangleq P\left(\theta,\left.\phi\right|\underline{w}^{ref}\right)$;
$P\left(\theta,\left.\phi\right|\underline{w}^{ref}\right)=\left|AF\left(\theta,\left.\phi\right|\underline{w}^{ref}\right)\right|^{2}${]}
fitting the upper $UM\left(\theta,\phi\right)$ and lower $LM\left(\theta,\phi\right)$
mask constraints, determine (}\textcolor{black}{\emph{a}}\textcolor{black}{)
the set of the minimum-norm/radiating weighting coefficients $\underline{w}^{RA}$,
which radiate a} \textcolor{black}{\emph{MN}} \textcolor{black}{power
pattern $P^{RA}\left(\theta,\phi\right)$ {[}$P^{RA}\left(\theta,\phi\right)\triangleq P\left(\theta,\left.\phi\right|\underline{w}^{RA}\right)$;
$P\left(\theta,\left.\phi\right|\underline{w}^{RA}\right)=\left|AF\left(\theta,\left.\phi\right|\underline{w}^{RA}\right)\right|^{2}${]}
close to the reference one within a certain tolerance $\xi$\begin{equation}
\frac{\int_{0}^{2\pi}\int_{0}^{\frac{\pi}{2}}\left|P^{RA}\left(\theta,\phi\right)-P^{ref}\left(\theta,\phi\right)\right|\sin\theta d\theta d\phi}{\int_{0}^{2\pi}\int_{0}^{\frac{\pi}{2}}P^{ref}\left(\theta,\phi\right)\sin\theta d\theta d\phi}\leq\xi;\label{eq:_MN.power.pattern.matching}\end{equation}
(}\textcolor{black}{\emph{b}}\textcolor{black}{) the set of non-radiating
coefficients $\underline{w}^{NR}$ so that additional user-defined
geometrical and/or electrical constraints, coded into a dedicated
objective function $\Phi_{C}\left(\underline{w}\right)$ to be minimized,
are fulfilled.}
\end{quotation}

\section{\textcolor{black}{\emph{NR}} \textcolor{black}{Current Based Constrained
Array Synthesis (}\textcolor{black}{\emph{NR-CAS}}\textcolor{black}{)
Method \label{sec:Non-Radiating-Current-Based}}}

\noindent \textcolor{black}{In order to exploit the non-uniqueness
of the} \textcolor{black}{\emph{IS}} \textcolor{black}{problem at
hand and the existence of the} \textcolor{black}{\emph{NR}} \textcolor{black}{currents
as additional} \textcolor{black}{\emph{DoF}}\textcolor{black}{s, the}
\textcolor{black}{\emph{CAS}} \textcolor{black}{problem is numerically
solved through the procedure sketched in Fig. 2 and detailed in the
following:}

\begin{itemize}
\item \textcolor{black}{\emph{Step}} \textcolor{black}{1} \textcolor{black}{\emph{-}}
\textbf{\textcolor{black}{\emph{EM Field Discretization}}}\textcolor{black}{.
Discretize the angular domain into $M$ samples ($M\geq N$) and fill
the vector $\underline{AF}$ whose $m$-th ($m=1,...,M$) entry is
$AF\left(\theta_{m},\phi_{m}\right)$, compute the radiation operator
matrix $\mathcal{G}=\left\{ g_{mn};,\, m=1,...,M;\, n=1,...,N\right\} $
of dimensions $M\times N$, whose ($m$, $n$)-th ($m=1,...,M$; $n=1,...,N$)
complex-valued entry is\begin{equation}
g_{mn}=e^{j\frac{2\pi}{\lambda}\left[\left(x_{n}\sin\theta_{m}\cos\phi_{m}+y_{n}\sin\theta_{m}\sin\phi_{m}\right)\right]},\label{eq:_steering.vector}\end{equation}
to rewrite (\ref{eq:_array.factor}) in matrix form, that is $\underline{AF}=\mathcal{G}\underline{w}$;}
\item \textcolor{black}{\emph{Step}} \textcolor{black}{2} \textcolor{black}{\emph{-}}
\textbf{\textcolor{black}{\emph{SVD of Radiation Operator}}}\textcolor{black}{.
Perform the singular value decomposition (}\textcolor{black}{\emph{SVD}}\textcolor{black}{)
of the radiation operator matrix, $\mathcal{G}$,\begin{equation}
\mathcal{G}=\mathcal{U}\Sigma\mathcal{V}^{*},\label{eq:_SVD}\end{equation}
$\mathcal{U}$ and $\mathcal{V}$ being the matrices whose $m$-th
($m=1,...,M$) and $n$-th ($n=1,...,N$) columns are the left, $\underline{u}_{m}$,
and the right, $\underline{v}_{n}$, singular vectors, respectively,
of unit norm (i.e., $\left\Vert \underline{u}_{m}\right\Vert _{2}=1$,
$\forall m$; $\left\Vert \underline{v}_{n}\right\Vert _{2}=1$, $\forall n$)
and orthogonal (i.e., $\left\langle \underline{u}_{m},\underline{u}_{l}\right\rangle =0$,
$\forall m$, $l=1,...,M$, $m\neq l$; $\left\langle \underline{v}_{n},\underline{v}_{l}\right\rangle =0$,
$\forall n$, $l=1,...,N$, $n\neq l$, $\left\langle \cdot,\cdot\right\rangle $
being the scalar product). In (\ref{eq:_SVD}), $\Sigma=\textrm{diag}\left\{ \sigma_{n};\, n=1...,N\right\} $
is the diagonal matrix whose non-null entries are the $N$ singular
values listed in descending order (i.e., $\sigma_{n}\geq\sigma_{n+1}$,
$n=1,...,N-1$) and $.^{*}$ stands for conjugate transpose;}
\item \textcolor{black}{\emph{Step}} \textcolor{black}{3 -} \textbf{\textcolor{black}{\emph{MN
Excitations}}} \textbf{\textcolor{black}{}}\textbf{\textcolor{black}{\emph{Definition}}}\textcolor{black}{.
Compute the radiating coefficients as follows\begin{equation}
\underline{w}^{RA}=\sum_{s=1}^{S}\frac{1}{\sigma_{s}}\left\langle \underline{u}_{s}^{*},\underline{AF}^{ref}\right\rangle \underline{v}_{s}\label{eq:_MN-excitations}\end{equation}
where $S=\arg\max_{n=1,...,N}\left\{ \hat{\sigma}_{n}>\chi\right\} $
($\hat{\sigma}_{n}\triangleq\frac{\sigma_{n}}{\sigma_{1}}$, $S\leq N$)
is the number of normalized singular values above a user-defined threshold
$\chi$ so that (\ref{eq:_MN.power.pattern.matching}) holds true;}
\item \noindent \textcolor{black}{\emph{Step}} \textcolor{black}{4} \textcolor{black}{\emph{-}}
\textbf{\textcolor{black}{\emph{NR Excitations Optimization}}}\textcolor{black}{.
Compute the set of the} \textcolor{black}{\emph{non-radiating}} \textcolor{black}{excitations
as \begin{equation}
\underline{w}^{NR}=\sum_{q=1}^{N-S}\gamma_{q}\underline{v}_{S+q}\label{eq:_NR-excitations}\end{equation}
where $\gamma_{q}$ is the $q$-th ($q=1,...,N-S$)} \textcolor{black}{\emph{NR}}
\textcolor{black}{expansion coefficient determined through the optimization
of a user-defined cost function, $\Phi_{C}\left(\underline{w}\right)=\Phi_{C}\left(\underline{w}^{RA}+\underline{w}^{NR}\right)$,
that mathematically codes the additional user-defined {}``}\textcolor{black}{\emph{non
functional}}\textcolor{black}{'' requirements on the element excitations;}
\item \textcolor{black}{\emph{Step}} \textcolor{black}{5 -} \textbf{\textcolor{black}{\emph{CAS
Problem Solution}}}\textcolor{black}{. Determine the excitations of
the array elements, $\underline{w}$, to be implemented in the} \textcolor{black}{\emph{BFN}}
\textcolor{black}{(see} \textcolor{black}{\emph{Appendix}}\textcolor{black}{)
by substituting (\ref{eq:_MN-excitations}) and (\ref{eq:_NR-excitations})
in (\ref{eq:_excitation.MN-plus-NR}).}
\end{itemize}

\section{\textcolor{black}{Numerical Results \label{sec:Numerical-Results}}}

\noindent \textcolor{black}{This section is devoted to illustrate
the features of the proposed} \textcolor{black}{\emph{NR-CAS}} \textcolor{black}{method
and to show its performance on a set of representative} \textcolor{black}{\emph{CAS}}
\textcolor{black}{problems. Both linear (1}\textcolor{black}{\emph{D}}\textcolor{black}{)
and planar (2}\textcolor{black}{\emph{D}}\textcolor{black}{)} \textcolor{black}{\emph{PA}}\textcolor{black}{s
as well as different {}``}\textcolor{black}{\emph{functional}}\textcolor{black}{''
design goals and user-defined {}``}\textcolor{black}{\emph{non-functional}}\textcolor{black}{''
constraints have been addressed. More specifically, the following
test cases (}\textcolor{black}{\emph{TC}}\textcolor{black}{s) have
been dealt with:}

\begin{itemize}
\item \noindent \textcolor{black}{the minimization of the dynamic range
ratio (}\textcolor{black}{\emph{DRR}}\textcolor{black}{) (}\textcolor{black}{\emph{TC}}\textcolor{black}{1);}
\item \noindent \textcolor{black}{the presence of a {}``}\textcolor{black}{\emph{forbidden}}\textcolor{black}{''
region within the antenna aperture where no elementary radiating elements
can be placed (}\textcolor{black}{\emph{TC}}\textcolor{black}{2);}
\item \noindent \textcolor{black}{the use of amplifiers with a limited number
of bits, $B$, thus a reduced set of amplitude values for the element
excitations of the} \textcolor{black}{\emph{PA}} \textcolor{black}{(}\textcolor{black}{\emph{TC}}\textcolor{black}{3).}
\end{itemize}
\noindent \textcolor{black}{The optimization of $\Phi_{C}\left(\underline{w}\right)$
to determine the set of complex} \textcolor{black}{\emph{NR}} \textcolor{black}{coefficients
has been performed with a suitable customization of the Particle Swarm
Optimizer (}\textcolor{black}{\emph{PSO}}\textcolor{black}{) \cite{Robinson 2004}
by using the following calibration setup \cite{Rocca 2009a}: $T=N-S$
($T$ being the swarm dimension), $\omega=0.4$ ($\omega$ being the
inertial weight), $C_{1}=C_{2}=2.0$ ($C_{1}$ and $C_{2}$ being
the cognitive and the social acceleration coefficients, respectively).
The initial swarm has been randomly initialized also including the
null solution, namely setting $\gamma_{q}=0$ ($q=1,...,N-S$) in
(\ref{eq:_NR-excitations}), and all simulations have been performed
by running a non-optimized software code on a standard laptop PC at
2.4 GHz CPU with 2 GB of RAM.}

\noindent \textcolor{black}{In the first synthesis problem (}\textcolor{black}{\emph{TC}}\textcolor{black}{1),
the array architecture is a linear array of $N=32$ elements located
along the $y$-axis with a uniform inter-element spacing equal to
$d=0.3\lambda$ {[}i.e., $x_{n}=0$, $y_{n}=\left(n-1\right)\times d$
($n=1,...,N$){]}. The reference power pattern, which is only function
of $\theta$ due to the 1}\textcolor{black}{\emph{D}} \textcolor{black}{layout
{[}i.e., $P^{ref}\left(\theta,\phi\right)\rightarrow P^{ref}\left(\theta\right)${]},
has been chosen equal to an asymmetric shaped beam, namely a cosecant-squared
power pattern {[}Figs. 3(}\textcolor{black}{\emph{e}}\textcolor{black}{){]},
to fulfill a mask pattern characterized by the following features:
$SLL^{ref}=-20$ {[}dB{]} ($SLL$ being the sidelobe level), $RPE=1.0$
{[}dB{]} ($RPE$ being the main lobe ripple), and $FNBW=68$ {[}deg{]}
($FNBW$ being the first null beamwidth). The corresponding amplitude,
\{$\alpha_{n}^{ref}$; $n=1,...,N$\}, and phase, \{$\beta_{n}^{ref}$;
$n=1,...,N$\}, distributions of the array excitations in Figs. 3(}\textcolor{black}{\emph{a}}\textcolor{black}{)-3(}\textcolor{black}{\emph{b}}\textcolor{black}{)
have been computed with a linear programming (}\textcolor{black}{\emph{LP}}\textcolor{black}{)
approach \cite{Isernia 1998}. According to the proposed} \textcolor{black}{\emph{NR-CAS}}
\textcolor{black}{method, the synthesis of the array excitation set,
$\underline{w}$, starts with the definition of the} \textcolor{black}{\emph{MN}}
\textcolor{black}{excitation weights, $\underline{w}^{RA}$. Towards
this end, the} \textcolor{black}{\emph{SVD}} \textcolor{black}{of
the linear operator $\mathcal{G}$ has been carried out according
to (\ref{eq:_SVD}) for determining the singular values, \{$\sigma_{n}$;
$n=1,...,N$\}, and the corresponding sets of eigenvectors, $\mathcal{U}=\left\{ \underline{u}_{m};\, m=1,...,M\right\} $
and $\mathcal{V}=\left\{ \underline{v}_{n};\, n=1...,N\right\} $.
Figure 4(}\textcolor{black}{\emph{a}}\textcolor{black}{) plots the
normalized singular values, \{$\hat{\sigma}_{n}$; $n=1,...,N$\},
that exhibit the well-known {}``knee'' trend. The magnitude and
the phase of the} \textcolor{black}{\emph{MN}} \textcolor{black}{excitations,
$\underline{w}^{RA}$, computed by setting different values of the}
\textcolor{black}{\emph{SVD}} \textcolor{black}{truncation threshold,
$\chi$, are reported in Figs. 4(}\textcolor{black}{\emph{c}}\textcolor{black}{)-4(}\textcolor{black}{\emph{d}}\textcolor{black}{),
while the corresponding power patterns, $\left.P^{RA}\left(\theta\right)\right\rfloor _{\chi}$,
together with the reference one, $P^{ref}\left(\theta\right)$, are
shown in Fig. 4(}\textcolor{black}{\emph{b}}\textcolor{black}{), respectively.
As expected, lower values of $\chi$ imply a better matching between
$P^{RA}\left(\theta\right)$ and $P^{ref}\left(\theta\right)$. When
choosing $\chi=\chi_{2}=3.5\times10^{-3}$ {[}Fig. 4(}\textcolor{black}{\emph{a}}\textcolor{black}{){]},
the number of singular values above the threshold $\chi$ turns out
to be $S=24$ and the pattern tolerance (\ref{eq:_MN.power.pattern.matching})
is $\xi<10^{-4}$. The} \textcolor{black}{\emph{NR}} \textcolor{black}{coefficients,
\{$\gamma_{q}$; $q=1,...,N-S$\}, have been then synthesized by means
of the} \textcolor{black}{\emph{PSO}}\textcolor{black}{-based optimization
strategy through the minimization of the} \textcolor{black}{\emph{DRR}}
\textcolor{black}{thus\begin{equation}
\left.\Phi_{C}\left(\underline{w}\right)\right\rfloor _{TC1}\triangleq\frac{\max_{n=1,...,N}\left\{ \alpha_{n}\right\} }{\min_{n=1,...,N}\left\{ \alpha_{n}\right\} }\,.\label{eq:_cost-function.TC1_DRR}\end{equation}
The magnitude and the phase of the} \textcolor{black}{\emph{NR}} \textcolor{black}{coefficients
in (\ref{eq:_NR-excitations}), \{$\gamma_{q}$; $q=1,...,N-S$\},
of the best trial solution of the swarm at different representative
iterations until the convergence ($i=I$, $I=500$) are shown in Fig.
3(}\textcolor{black}{\emph{c}}\textcolor{black}{) and Fig. 3(}\textcolor{black}{\emph{d}}\textcolor{black}{),
respectively. The corresponding element-level excitations, \{$w_{n}=\left.w_{n}^{RA}\left(\chi\right)\right\rfloor _{\chi=\chi_{2}}+w_{n}^{NR}$;
$n=1...,N$\}, and the arising power patterns are given in Figs. 3(}\textcolor{black}{\emph{a}}\textcolor{black}{)-3(}\textcolor{black}{\emph{b}}\textcolor{black}{)
and in Fig. 3(}\textcolor{black}{\emph{e}}\textcolor{black}{), respectively.
For the sake of completeness, the evolution of the cost function value
of the best solution of the swarm, $\Phi_{i}^{opt}$ ($\Phi_{i}^{opt}\triangleq\min_{t=1,...,T}\left\{ \Phi_{C}\left(\left[\underline{w}\right]_{t}^{i}\right)\right\} $,
$\left[\underline{w}\right]_{t}^{i}$ being the $t$-th trial excitation
set at the $i$-th iteration), versus the iteration index $i$ ($i=0,...,I$)
is reported in Fig. 5. As expected, the optimized} \textcolor{black}{\emph{NR-CAS}}
\textcolor{black}{solution outperforms the minimum-norm (labeled as}
\textcolor{black}{\emph{RA}}\textcolor{black}{) one in terms of the
design goal ($DRR^{RA}\approx31.55$ vs. $DRR^{NR-CAS}\approx3.40$),
while the value of the quality factor \cite{Hansen 2006}\begin{equation}
Q=\frac{\sum_{n=1}^{N}\alpha_{n}^{2}}{\int_{0}^{\pi}\int_{0}^{2\pi}P\left(\theta,\phi\right)\sin\theta d\theta d\phi}\label{eq:_Q-factor}\end{equation}
slightly increases ($Q^{RA}=0.61$ vs. $Q^{NR-CAS}=0.75$).}

\noindent \textcolor{black}{In order to assess the reliability of
the synthesized} \textcolor{black}{\emph{NR-CAS}} \textcolor{black}{solution,
a more realistic model for the elementary radiator, instead of an
ideal one, has been considered. More specifically, the single-polarization
aperture-coupled stacked square patch} \textcolor{black}{\emph{}}\textcolor{black}{antenna
\cite{Targonski 1998} in Figs. 6(}\textcolor{black}{\emph{a}}\textcolor{black}{)-6(}\textcolor{black}{\emph{b}}\textcolor{black}{)
and Tab. I, which resonates at $f=3.5$ {[}GHz{]}, has been chosen.
In particular, the} \textcolor{black}{\emph{}}\textcolor{black}{antenna
consists of two stacked microstrip patch radiators printed on top
of two substrates and it is fed by a microstrip line coupled through
a rectangular slot. The slot is etched in a ground-plane that separates
the patches and the feeding line. The dielectric material of the substrates
is Rogers {}``RT/Duroid 3003'' \cite{Rogers 2022}, which is characterized
by a relative permittivity and a loss tangent equal to $\varepsilon_{r}=3.0$
and $\tan\delta=0.0016$, respectively. The power pattern of the array
layout in Fig. 6(}\textcolor{black}{\emph{c}}\textcolor{black}{) has
been simulated with the finite-element full-wave solver} \textcolor{black}{\emph{Ansys
HFSS}} \textcolor{black}{\cite{HFSS 2021} to properly take into account
mutual coupling effects. Figure 7 compares the ideal} \textcolor{black}{\emph{NR-CAS}}
\textcolor{black}{pattern and the numerically-simulated one by showing
that the main deviations between the two plots only arise in the region
close to end-fire (i.e., $\sin\theta=1$), while there is a good match
in the overall.}

\noindent \textcolor{black}{In the second test case (}\textcolor{black}{\emph{TC}}\textcolor{black}{2),
the goal is to exploit the} \textcolor{black}{\emph{NR}} \textcolor{black}{current
components to create a {}``}\textcolor{black}{\emph{forbidden}}\textcolor{black}{{}``
region, $\Psi$, within the array aperture where no array elements
can be placed (Fig. 8). Towards this end, the objective cost function
$\Phi_{C}\left(\underline{w}\right)$ has been defined as follows\begin{equation}
\left.\Phi_{C}\left(\underline{w}\right)\right\rfloor _{TC2}\triangleq\sum_{\left(x_{n},y_{n}\right)\in\Psi,\,\forall n}\alpha_{n}\label{eq:_cost-function.TC2_Forbidden-region}\end{equation}
and the benchmark example was that of a square planar array having
$N=16\times16$ elements located on a regular square lattice with
uniform inter-element spacing equal to $d=0.45\lambda$ along the
$x$ and $y$ direction. The reference power pattern is a flat-top
power pattern in Fig. 9(}\textcolor{black}{\emph{e}}\textcolor{black}{)
with asymmetric sidelobes that fits the lower and upper masks reported
in Fig. 9(}\textcolor{black}{\emph{a}}\textcolor{black}{) and Fig.
9(}\textcolor{black}{\emph{b}}\textcolor{black}{), respectively, featuring
$RPE=0.5$ {[}dB{]} and $FNBW=50$ {[}deg{]} along both the $x$-axis
and $y$-axis. The reference amplitude, \{$\alpha_{n}^{ref}$; $n=1,...,N$\},
and phase, \{$\beta_{n}^{ref}$; $n=1,...,N$\}, excitations in Fig.
9(}\textcolor{black}{\emph{c}}\textcolor{black}{) and Fig. 9(}\textcolor{black}{\emph{d}}\textcolor{black}{),
respectively, have been again obtained through} \textcolor{black}{\emph{LP}}\textcolor{black}{.
Once performed the} \textcolor{black}{\emph{SVD}} \textcolor{black}{of
$\mathcal{G}$, the distribution of the normalized singular values,
\{$\hat{\sigma}_{n}$; $n=1,...,N$\}, is that in Fig. 10. To achieve
a value of the power pattern matching error (\ref{eq:_MN.power.pattern.matching})
$\xi<10^{-4}$, the threshold $\chi$ has been set to $\chi=7.2\times10^{-3}$
so that the number of singular values turned out equal to $S=236$.
The synthesis of the} \textcolor{black}{\emph{NR}} \textcolor{black}{coefficients,
\{$\gamma_{q}$; $q=1,...,N-S$\}, has been yielded by running the}
\textcolor{black}{\emph{PSO}}\textcolor{black}{-based optimization
for $I=2000$ iterations due to the larger number of unknowns} \textbf{\textcolor{black}{}}\textcolor{black}{as
compared to the previous example. The behavior of the optimal value
of the cost function (\ref{eq:_cost-function.TC2_Forbidden-region}),
$\Phi_{i}^{opt}$, during the iterative minimization is given in Fig.
11, while the magnitudes, \{$\left|\gamma_{q}\right|$; $q=1,...,N-S$\},
and the phases, \{$\angle\gamma_{q}$; $q=1,...,N-S$\}, of the} \textcolor{black}{\emph{NR}}
\textcolor{black}{coefficients of $\left[\underline{w}\right]_{opt}^{i}$
($\left[\underline{w}\right]_{opt}^{i}$ being the best solution of
the swarm at the $i$-th ($i=0,...,I$) iteration, $\left[\underline{w}\right]_{opt}^{i}\triangleq\arg\left[\min_{t=1,...,T}\left\{ \Phi_{C}\left(\left[\underline{w}\right]_{t}^{i}\right)\right\} \right]$)
at both the iterations $i=\left\{ 0,\,500,\,1000\right\} $ and convergence
($i=I^{conv}=1895$ since $\left.\Phi_{i}^{opt}\right\rfloor _{i=I^{conv}}=0$)
are shown in Fig. 12(}\textcolor{black}{\emph{a}}\textcolor{black}{)
and Fig. 12(}\textcolor{black}{\emph{b}}\textcolor{black}{), respectively.}

\noindent \textcolor{black}{To assess the fulfillment of the {}``}\textcolor{black}{\emph{forbidden}}\textcolor{black}{''
region constraint, let us analyze the magnitude, \{$\alpha_{n}$;
$n=1,...,N$\}, and the phase, \{$\beta_{n}$; $n=1,...,N$\}, of
the corresponding element-level excitations, \{$w_{n}$; $n=1...,N$\},
in Figs. 13(}\textcolor{black}{\emph{a}}\textcolor{black}{)-13(}\textcolor{black}{\emph{d}}\textcolor{black}{)
and Figs. 13(}\textcolor{black}{\emph{e}}\textcolor{black}{)-13(}\textcolor{black}{\emph{h}}\textcolor{black}{),
respectively. As expected, starting from the reference distribution
{[}$i=0$ - Fig. 13(}\textcolor{black}{\emph{a}}\textcolor{black}{){]}
that violates the null-excitations condition in the whole area $\Psi$,
the solution improves during the iterations {[}Figs. 13(}\textcolor{black}{\emph{b}}\textcolor{black}{)-13(}\textcolor{black}{\emph{c}}\textcolor{black}{){]}
until the convergence {[}Fig. 13(}\textcolor{black}{\emph{d}}\textcolor{black}{){]}
when $\Phi_{C}\left(\left[\underline{w}\right]_{opt}^{I^{conv}}\right)=0$.
On the other hand, the excitations profile in Fig. 13(}\textcolor{black}{\emph{d}}\textcolor{black}{)
and Fig. 13(}\textcolor{black}{\emph{h}}\textcolor{black}{) radiates
a power pattern whose cuts along the principal planes, $\phi=0$ {[}deg{]}
and $\phi=90$ {[}deg{]}, are reported in Fig. 12(}\textcolor{black}{\emph{c}}\textcolor{black}{)
and Fig. 12(}\textcolor{black}{\emph{d}}\textcolor{black}{), respectively,
to confirm the mask-matching of the} \textcolor{black}{\emph{NR-CAS}}
\textcolor{black}{solution.}

\noindent \textcolor{black}{Analogously to the test case} \textcolor{black}{\emph{TC}}\textcolor{black}{1,
the reliability of the synthesis process has been checked also here
by considering the same radiator {[}Figs. 6(}\textcolor{black}{\emph{a}}\textcolor{black}{)-6(}\textcolor{black}{\emph{b}}\textcolor{black}{){]}
to simulate with} \textcolor{black}{\emph{Ansys HFSS}} \textcolor{black}{\cite{HFSS 2021}
the non-ideal planar array in Fig. 14. The comparison between the
ideal and the full-wave} \textcolor{black}{\emph{NR-CAS}} \textcolor{black}{patterns
along $\phi=0$ {[}deg{]} {[}Fig. 15(}\textcolor{black}{\emph{a}}\textcolor{black}{){]}
and $\phi=90$ {[}deg{]} {[}Fig. 15(}\textcolor{black}{\emph{b}}\textcolor{black}{){]}
highlights that also in this case, the main deviations from the ideal
behavior occur far from broadside and within the sidelobe region.}

\noindent \textcolor{black}{The last test case (}\textcolor{black}{\emph{TC}}\textcolor{black}{3)
is concerned with the use of the proposed} \textcolor{black}{\emph{NR}}\textcolor{black}{-based
method to simplify the} \textcolor{black}{\emph{BFN}}\textcolor{black}{.
Accordingly, digital amplifiers with $B=2$ bits have been considered
so that only four (i.e., $\left.2^{B}\right\rfloor _{B=2}=4$) possible
amplitude values were available: $\underline{\alpha}^{trg}=\left\{ 0.25,\,0.50,\,0.75,\,1.00\right\} $.
As for the array, the layout of the previous example (}\textcolor{black}{\emph{TC}}\textcolor{black}{2)
has been kept, but setting different lower and upper masks for the
reference power {[}Figs. 16(}\textcolor{black}{\emph{a}}\textcolor{black}{)-16(}\textcolor{black}{\emph{b}}\textcolor{black}{){]}
with $RPE=1.0$ {[}dB{]} and $FNBW=45$ {[}deg{]} in both Cartesian
planar axes. Therefore, the} \textcolor{black}{\emph{LP}}\textcolor{black}{-computed
excitations are those in Figs. 16(}\textcolor{black}{\emph{c}}\textcolor{black}{)-16(}\textcolor{black}{\emph{d}}\textcolor{black}{),
while the corresponding reference flat-top power pattern is shown
in Fig. 16(}\textcolor{black}{\emph{e}}\textcolor{black}{). Since
the array geometry is the same of} \textcolor{black}{\emph{TC}}\textcolor{black}{2,
the entries of $\mathcal{G}$ (\ref{eq:_steering.vector}) do not
change as well as the normalized singular values, \{$\sigma_{n}$;
$n=1,...,N$\} (Fig. 10) and the eigenvectors, $\mathcal{U}=\left\{ \underline{u}_{m};\, m=1,...,M\right\} $
and $\mathcal{V}=\left\{ \underline{v}_{n};\, n=1...,N\right\} $.
Moreover, the same values of $\chi$ and $S$ (i.e., $\chi=7.2\times10^{-3}$
and $S=236$) have been used, as well. The minimization of the cost
function to enforce quantized amplitude excitations\begin{equation}
\left.\Phi_{C}\left(\underline{w}\right)\right\rfloor _{TC3}=\sum_{n=1}^{N}\min_{i=1,...,2^{B}}\left|\alpha_{n}-\alpha_{i}^{trg}\right|\label{eq:_cost-function.TC3_Quantized-amplitudes}\end{equation}
stops after $i=I^{conv}=1651$ iterations (Fig. 17). Figure 18 shows
the evolution of both the} \textcolor{black}{\emph{NR}} \textcolor{black}{coefficients
{[}Figs. 18(}\textcolor{black}{\emph{a}}\textcolor{black}{)-18(}\textcolor{black}{\emph{b}}\textcolor{black}{){]}
of $\left[\underline{w}\right]_{opt}^{i}$ and the radiated patterns
{[}Figs. 18(}\textcolor{black}{\emph{c}}\textcolor{black}{)-18(}\textcolor{black}{\emph{d}}\textcolor{black}{){]}
during the iterative optimization, while the corresponding element-level
excitations, \{$w_{n}$; $n=1...,N$\}, are reported in Fig. 19 in
terms of magnitude (\ref{eq:_W Rossiglione}), \{$\alpha_{n}$; $n=1,...,N$\}
{[}Figs. 19(}\textcolor{black}{\emph{a}}\textcolor{black}{)-19(}\textcolor{black}{\emph{d}}\textcolor{black}{){]},
and phase (\ref{eq:_Bleggio Rocca's Land}), \{$\beta_{n}$; $n=1,...,N$\}
{[}Figs. 19(}\textcolor{black}{\emph{e}}\textcolor{black}{)-19(}\textcolor{black}{\emph{h}}\textcolor{black}{){]},
coefficients. By observing the magnitude terms on the left column
of Fig. 19, one can notice that thanks to the} \textcolor{black}{\emph{NR}}
\textcolor{black}{contribution, \{$w_{n}^{NR}$; $n=1...,N$\}, the
convergence ($i=I^{conv}=1651$) amplitudes {[}Fig. 19(}\textcolor{black}{\emph{d}}\textcolor{black}{){]}
are quantized to the target values {[}i.e., $\alpha_{n}\in\left\{ 0.25,\,0.50,\,0.75,\,1.00\right\} $
($n=1,...,N$){]}. Of course, this does not impact on the pattern
radiated by the array as confirmed by the pictures of the principal
cuts of the power pattern in Figs. 18(}\textcolor{black}{\emph{c}}\textcolor{black}{)-18(}\textcolor{black}{\emph{d}}\textcolor{black}{).
Lastly, the good agreement with the prediction given by the full-wave
simulator (Fig. 20) further assesses the reliability of the} \textcolor{black}{\emph{NR-CAS}}
\textcolor{black}{synthesis.}

\section{\noindent \textcolor{black}{Conclusions and Final Remarks \label{sec:Conclusions}}}

\noindent \textcolor{black}{An innovative method, namely the} \textcolor{black}{\emph{NR-CAS}}
\textcolor{black}{approach, for the constrained synthesis of} \textcolor{black}{\emph{PA}}
\textcolor{black}{antennas has been proposed. It is aimed at synthesizing
the complex (amplitude and} phase) excitations of the array elements
to fulfill {}``\emph{non-functional}'' constraints, while fitting
desired power pattern masks. More specifically, the array weights
are expressed as the linear combination of a radiating minimum-norm
term, to enforce the {}``\emph{functional}'' far-field behavior,
and a \emph{NR} component that generates a null \emph{EM} field outside
the array aperture, but allows one to properly address other requirements
on the array structure. The synthesis of the \emph{NR} term is yielded
by determining its expansion coefficients with respect to a set of
basis functions derived from the \emph{SVD} of the radiation operator.
Towards this end, an optimization problem is formulated by defining
a suitable cost function, which mathematically quantifies the degree
of fulfillment of the {}``\emph{non-functional}'' constraints, to
be minimized according to a \emph{PSO}-based iterative process.

\noindent From the numerical assessment, the following main outcomes
can be drawn:

\begin{itemize}
\item the \emph{NR-CAS} method is a versatile technique able to properly
handle the constrained synthesis of the \emph{PA} excitations such
as the optimization of the \emph{DRR}, the inclusion of a {}``\emph{forbidden}''
region within the array aperture, and the use of digital amplifiers
with only few amplitude levels;
\item the \emph{NR-CAS} method is computationally very efficient since the
synthesis is concerned with a number of \emph{DoF}s {[}i.e., the ($N-S$)
\emph{NR} expansion coefficients{]} much smaller than those of a standard
design, which is equal to the number of antenna array elements, the
\emph{DoF}s being the whole set of $N$ array excitations.
\end{itemize}
Future research activities, beyond the scope of this work, will be
aimed at extending the proposed \emph{NR-CAS} method to other array
geometries and/or architectures (e.g., unconventional arrays \cite{Rocca 2016})
as well as different types of {}``\emph{non-functional}'' constraints
or requirements.

\section*{Appendix}

The complex excitation, $w_{n}$ ($w_{n}=\alpha_{n}e^{j\beta_{n}}$),
of the $n$-th ($n=1,...,N$) radiating element of the \emph{PA},
is obtained from the corresponding $n$-th ($n=1,...,N$) \emph{radiating},
$w_{n}^{RA}$ ($w_{n}^{RA}=\alpha_{n}^{RA}e^{j\beta_{n}^{RA}}$),
and \emph{non-radiating}, $w_{n}^{NR}$ ($w_{n}^{NR}=\alpha_{n}^{NR}e^{j\beta_{n}^{NR}}$),
components. More specifically, the $n$-th ($n=1,...,N$) amplitude
coefficient $\alpha_{n}$ is given by\begin{equation}
\alpha_{n}=\sqrt{\left(\alpha_{n}^{RA}\right)^{2}+2\alpha_{n}^{NR}\alpha_{n}^{RA}\left[\cos\left(\beta_{n}^{RA}-\beta_{n}^{NR}\right)\right]+\left(\alpha_{n}^{NR}\right)^{2}},\label{eq:_W Rossiglione}\end{equation}
while the $n$-th ($n=1,...,N$) phase term $\beta_{n}$ is yielded
as follows\begin{equation}
\beta_{n}=\arctan\left(\frac{\alpha_{n}^{RA}\sin\beta_{n}^{RA}+\alpha_{n}^{NR}\sin\beta_{n}^{NR}}{\alpha_{n}^{RA}\cos\left(\beta_{n}^{RA}\right)+\alpha_{n}^{NR}\cos\left(\beta_{n}^{NR}\right)}\right).\label{eq:_Bleggio Rocca's Land}\end{equation}

\section*{\noindent Acknowledgements}

\noindent This work benefited from the networking activities carried
out within \textbf{}the Project DICAM-EXC (Grant L232/2016) funded
by the Italian Ministry of Education, Universities and Research (MUR)
within the 'Departments of Excellence 2023-2027' Program (CUP: E63C22003880001).
Moreover, it benefited from the networking activities carried out
within \textbf{}the Project SEME@TN - Smart ElectroMagnetic Environment
in TrentiNo funded by the Autonomous Province of Trento (CUP: C63C22000720003),
the Project AURORA - Smart Materials for Ubiquitous Energy Harvesting,
Storage, and Delivery in Next Generation Sustainable Environments
funded by the Italian Ministry for Universities and Research within
the PRIN-PNRR 2022 Program (CUP: E53D23014760001), the Project National
Centre for HPC, Big Data and Quantum Computing (CN HPC) funded by
the European Union - NextGenerationEU within the PNRR Program (CUP:
E63C22000970007), the Project Telecommunications of the Future {[}PE00000001
- program RESTART, Project 6GWINET (CUP: D43C22003080001), Project
MOSS (CUP: J33C22002880001), Project IN (CUP: J33C22002880001), Project
EMS-MMDV (CUP: J33C22002880001), Project TRIBOLETTO (CUP: B53C22003970001),
and Project SMART (CUP: E63C22002040007){]}, funded by European Union
under the Italian National Recovery and Resilience Plan (NRRP) of
NextGenerationEU, and the support of the Natural Science Basic Research
Program of Shaanxi Province (Grants No. 2022-JC-33, No. 2023-GHZD-35,
and No. 2024JC-ZDXM-25). Views and opinions expressed are however
those of the author(s) only and do not necessarily reflect those of
the European Union or the European Research Council. Neither the European
Union nor the granting authority can be held responsible for them.
A. Massa wishes to thank E. Vico for her never-ending inspiration,
support, guidance, and help.

\newpage
\section*{FIGURE CAPTIONS}

\begin{itemize}
\item \textbf{Figure 1.} Sketch of a 2\emph{D} array geometry.
\item \textbf{Figure 2.} Flowchart of the \emph{NR-CAS} method.
\item \textbf{Figure 3.} \emph{TC}1 \emph{DRR Optimization} ($1D$ Array,
$N=32$, $d=0.3\lambda$, Cosecant-squared Pattern, $SLL^{ref}=-20$
{[}dB{]}, $RPE=1.0$ {[}dB{]}, $FNBW=68$ {[}deg{]}) - Plot of (\emph{a})(\emph{c})
the magnitude and (\emph{b})(\emph{d}) the phase of (\emph{a})(\emph{b})
the excitations of the array, \{$w_{n}$; $n=1...,N$\}, and (\emph{c})(\emph{d})
the \emph{NR} expansion coefficients, \{$\gamma_{q}$; $q=1,...,N-S$\},
and (\emph{e}) the corresponding power patterns, $P\left(\theta\right)$,
at different representative iterations ($i$ being the iteration index)
of the \emph{PSO}-based optimization process.
\item \textbf{Figure 4.} \emph{TC}1 \emph{DRR Optimization} ($1D$ Array,
$N=32$, $d=0.3\lambda$, Cosecant-squared Pattern, $SLL^{ref}=-20$
{[}dB{]}, $RPE=1.0$ {[}dB{]}, $FNBW=68$ {[}deg{]}) - Plot of (\emph{a})
the normalized singular values of $\mathcal{G}$, \{$\hat{\sigma}_{n}$;
$n=1,...,N$\}, (\emph{c}) the magnitude and (\emph{d}) the phase
of the minimum-norm element excitations, \{$w_{n}^{RA}$; $n=1...,N$\},
and (\emph{b}) the corresponding power patterns, $P\left(\theta\right)$,
in correspondence with different values of the \emph{SVD} truncation
threshold, $\chi$.
\item \textbf{Figure 5.} \emph{TC}1 \emph{DRR Optimization} ($1D$ Array,
$N=32$, $d=0.3\lambda$, Cosecant-squared Pattern, $SLL^{ref}=-20$
{[}dB{]}, $RPE=1.0$ {[}dB{]}, $FNBW=68$ {[}deg{]}) - Plot of the
evolution the cost function value of the best solution of the swarm,
$\Phi_{i}^{opt}$ ($\Phi_{i}^{opt}\triangleq\min_{t=1,...,T}\left\{ \Phi_{C}\left(\left[\underline{w}\right]_{t}^{i}\right)\right\} $,
$\left[\underline{w}\right]_{t}^{i}$ being the $t$-th trial excitation
set at the $i$-th iteration), versus the iteration index $i$ ($i=0,...,I$;
$I=500$).
\item \textbf{Figure 6.} \emph{TC}1 \emph{DRR Optimization} ($1D$ Array,
$N=32$, $d=0.3\lambda$, Cosecant-squared Pattern, $SLL^{ref}=-20$
{[}dB{]}, $RPE=1.0$ {[}dB{]}, $FNBW=68$ {[}deg{]}) - Sketch of (\emph{a})
the exploded view and (\emph{b}) the top view of the aperture-coupled
stacked square patch antenna together with (\emph{c}) the \emph{HFSS}
model of the array layout.
\item \textbf{Figure 7.} \emph{TC}1 \emph{DRR Optimization} ($1D$ Array,
$N=32$, $d=0.3\lambda$, Cosecant-squared Pattern, $SLL^{ref}=-20$
{[}dB{]}, $RPE=1.0$ {[}dB{]}, $FNBW=68$ {[}deg{]}) - Comparison
between the ideal \emph{NR-CAS} pattern and the \emph{HFSS} numerically-simulated
one.
\item \textbf{Figure 8.} \emph{TC}2 \emph{Forbidden Region} ($2D$ Array,
$N=16\times16$, $d=0.45\lambda$, Flat-top Pattern, Asymmetric Sidelobes,
$RPE=0.5$ {[}dB{]}, $FNBW=50$ {[}deg{]}) - Sketch of the array layout
with a {}``\emph{forbidden}'' region, $\Psi$.
\item \textbf{Figure 9.} \emph{TC}2 \emph{Forbidden Region} ($2D$ Array,
$N=16\times16$, $d=0.45\lambda$, Flat-top Pattern, Asymmetric Sidelobes,
$RPE=0.5$ {[}dB{]}, $FNBW=50$ {[}deg{]}) - Plot of (\emph{a}) the
lower, $LM\left(\theta,\phi\right)$, and (\emph{b}) the upper, $UM\left(\theta,\phi\right)$,
power pattern masks, (\emph{c}) the magnitude and (\emph{d}) the phase
of the array reference excitations, \{$w_{n}^{ref}$; $n=1...,N$\},
along with (\emph{e}) the corresponding power pattern, $P^{ref}\left(\theta,\phi\right)$.
\item \textbf{Figure 10.} \emph{TC}2 \emph{Forbidden Region} ($2D$ Array,
$N=16\times16$, $d=0.45\lambda$, Flat-top Pattern, Asymmetric Sidelobes,
$RPE=0.5$ {[}dB{]}, $FNBW=50$ {[}deg{]}) - Plot of the normalized
singular values of $\mathcal{G}$, \{$\hat{\sigma}_{n}$; $n=1,...,N$\}.
\item \textbf{Figure 11.} \emph{TC}2 \emph{Forbidden Region} ($2D$ Array,
$N=16\times16$, $d=0.45\lambda$, Flat-top Pattern, Asymmetric Sidelobes,
$RPE=0.5$ {[}dB{]}, $FNBW=50$ {[}deg{]}) - Plot of the evolution
the cost function value of the best solution of the swarm, $\Phi_{i}^{opt}$
($\Phi_{i}^{opt}\triangleq\min_{t=1,...,T}\left\{ \Phi_{C}\left(\left[\underline{w}\right]_{t}^{i}\right)\right\} $,
$\left[\underline{w}\right]_{t}^{i}$ being the $t$-th trial excitation
set at the $i$-th iteration), versus the iteration index $i$ ($i=0,...,I$;
$I=2000$; $I^{conv}=1895$).
\item \textbf{Figure 12.} \emph{TC}2 \emph{Forbidden Region} ($2D$ Array,
$N=16\times16$, $d=0.45\lambda$, Flat-top Pattern, Asymmetric Sidelobes,
$RPE=0.5$ {[}dB{]}, $FNBW=50$ {[}deg{]}) - Plot of (\emph{a}) the
magnitude and (\emph{b}) the phase of the \emph{NR} expansion coefficients,
\{$\gamma_{q}$; $q=1,...,N-S$\}, and (\emph{e}) the corresponding
cuts of the power pattern, $P\left(\theta,\phi\right)$, in the (\emph{c})
$\phi=0$ {[}deg{]} and (\emph{d}) $\phi=90$ {[}deg{]} planes at
different representative iterations ($i$ being the iteration index)
of the \emph{PSO}-based optimization process.
\item \textbf{Figure 13.} \emph{TC}2 \emph{Forbidden Region} ($2D$ Array,
$N=16\times16$, $d=0.45\lambda$, Flat-top Pattern, Asymmetric Sidelobes,
$RPE=0.5$ {[}dB{]}, $FNBW=50$ {[}deg{]}) - Plot of (\emph{a})-(\emph{d})
the magnitude and (\emph{e})-(\emph{h}) the phase of the element excitations,
\{$w_{n}$; $n=1...,N$\}, at the $i$-th iteration of the PSO-based
optimization process: (\emph{a})(\emph{e}) $i=0$, (\emph{b})(\emph{f})
$i=500$, (\emph{c})(\emph{g}) $i=1000$, and (\emph{d})(\emph{h})
$i=I^{conv}$ ($I^{conv}=1895$).
\item \textbf{Figure 14.} \emph{TC}2 \emph{Forbidden Region} ($2D$ Array,
$N=16\times16$, $d=0.45\lambda$, Flat-top Pattern, Asymmetric Sidelobes,
$RPE=0.5$ {[}dB{]}, $FNBW=50$ {[}deg{]}) - Sketch of the \emph{HFSS}
model of the array layout.
\item \textbf{Figure 15.} \emph{TC}2 \emph{Forbidden Region} ($2D$ Array,
$N=16\times16$, $d=0.45\lambda$, Flat-top Pattern, Asymmetric Sidelobes,
$RPE=0.5$ {[}dB{]}, $FNBW=50$ {[}deg{]}) - Comparison between the
ideal \emph{NR-CAS} pattern and the \emph{HFSS} numerically-simulated
one along the planes: (\emph{a}) $\phi=0$ {[}deg{]} and (\emph{b})
$\phi=90$ {[}deg{]}.
\item \textbf{Figure 16.} \emph{TC}3 \emph{Quantized Amplitudes} ($2D$
Array, $N=16\times16$, $d=0.45\lambda$, Flat-top Pattern, $SLL^{ref}=-20$
{[}dB{]}, $RPE=1.0$ {[}dB{]}, $FNBW=45$ {[}deg{]}) - Plot of (\emph{a})
the lower, $LM\left(\theta,\phi\right)$, and (\emph{b}) the upper,
$UM\left(\theta,\phi\right)$, power pattern masks, (\emph{c}) the
magnitude and (\emph{d}) the phase of the array reference excitations,
\{$w_{n}^{ref}$; $n=1...,N$\}, along with (\emph{e}) the corresponding
power pattern, $P^{ref}\left(\theta,\phi\right)$.
\item \textbf{Figure 17.} \emph{TC}3 \emph{Quantized Amplitudes} ($2D$
Array, $N=16\times16$, $d=0.45\lambda$, Flat-top Pattern, $SLL^{ref}=-20$
{[}dB{]}, $RPE=1.0$ {[}dB{]}, $FNBW=45$ {[}deg{]}) - Plot of the
evolution the cost function value of the best solution of the swarm,
$\Phi_{i}^{opt}$ ($\Phi_{i}^{opt}\triangleq\min_{t=1,...,T}\left\{ \Phi_{C}\left(\left[\underline{w}\right]_{t}^{i}\right)\right\} $,
$\left[\underline{w}\right]_{t}^{i}$ being the $t$-th trial excitation
set at the $i$-th iteration), versus the iteration index $i$ ($i=0,...,I$;
$I=2000$; $I^{conv}=1651$).
\item \textbf{Figure 18.} \emph{TC}3 \emph{Quantized Amplitudes} ($2D$
Array, $N=16\times16$, $d=0.45\lambda$, Flat-top Pattern, $SLL^{ref}=-20$
{[}dB{]}, $RPE=1.0$ {[}dB{]}, $FNBW=45$ {[}deg{]}) - Plot of (\emph{a})
the magnitude and (\emph{b}) the phase of the \emph{NR} expansion
coefficients, \{$\gamma_{q}$; $q=1,...,N-S$\}, and (\emph{e}) the
corresponding cuts of the power pattern, $P\left(\theta,\phi\right)$,
in the (\emph{c}) $\phi=0$ {[}deg{]} and (\emph{d}) $\phi=90$ {[}deg{]}
planes at different representative iterations ($i$ being the iteration
index) of the \emph{PSO}-based optimization process.
\item \textbf{Figure 19.} \emph{TC}3 \emph{Quantized Amplitudes} ($2D$
Array, $N=16\times16$, $d=0.45\lambda$, Flat-top Pattern, $SLL^{ref}=-20$
{[}dB{]}, $RPE=1.0$ {[}dB{]}, $FNBW=45$ {[}deg{]}) - Plot of (\emph{a})-(\emph{d})
the magnitude and (\emph{e})-(\emph{h}) the phase of the element excitations,
\{$w_{n}$; $n=1...,N$\}, at the $i$-th iteration of the \emph{PSO}-based
optimization process: (\emph{a})(\emph{e}) $i=0$, (\emph{b})(\emph{f})
$i=100$, (\emph{c})(\emph{g}) $i=500$, and (\emph{d})(\emph{h})
$i=I^{conv}$ ($I^{conv}=1651$).
\item \textbf{Figure 20.} \emph{TC}3 \emph{Quantized Amplitudes} ($2D$
Array, $N=16\times16$, $d=0.45\lambda$, Flat-top Pattern, $SLL^{ref}=-20$
{[}dB{]}, $RPE=1.0$ {[}dB{]}, $FNBW=45$ {[}deg{]}) - Comparison
between the ideal \emph{NR-CAS} pattern and the \emph{HFSS} numerically-simulated
one along the planes: (\emph{a}) $\phi=0$ {[}deg{]} and (\emph{b})
$\phi=90$ {[}deg{]}.
\end{itemize}

\section*{TABLE CAPTIONS}

\begin{itemize}
\item \textbf{Table I.} Aperture-coupled stacked square patch antenna descriptors.
\end{itemize}
\newpage
\begin{center}~\vfill\end{center}

\begin{center}\begin{tabular}{c}
\includegraphics[%
  width=0.90\columnwidth]{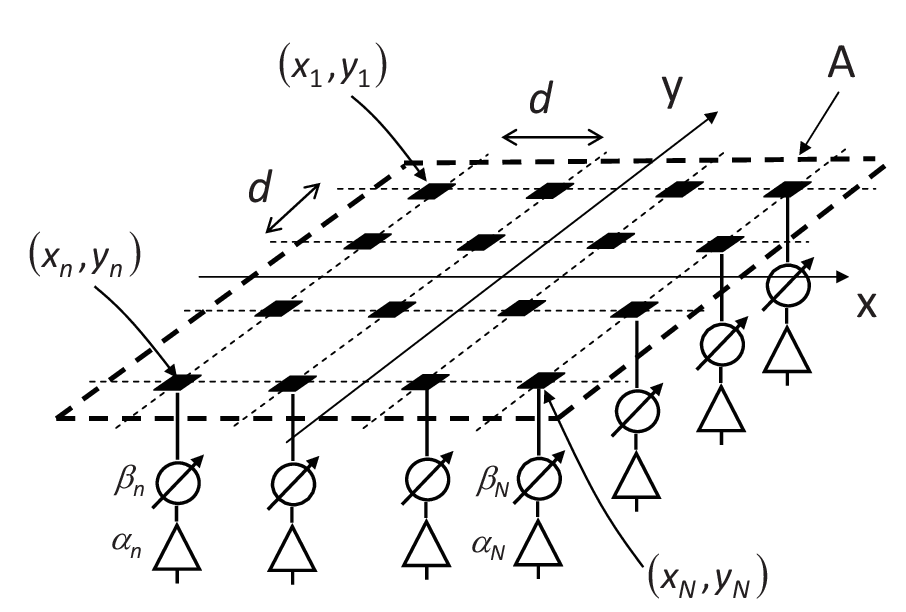}\tabularnewline
\end{tabular}\end{center}

\begin{center}~\vfill\end{center}

\begin{center}\textbf{Fig. 1 - Poli} \textbf{\emph{et al.,}} {}``Inverse
Source Method for Constrained Phased Array Synthesis ...''\end{center}

\newpage
\begin{center}~\vfill\end{center}

\begin{center}\begin{tabular}{c}
\includegraphics[%
  width=1.0\columnwidth]{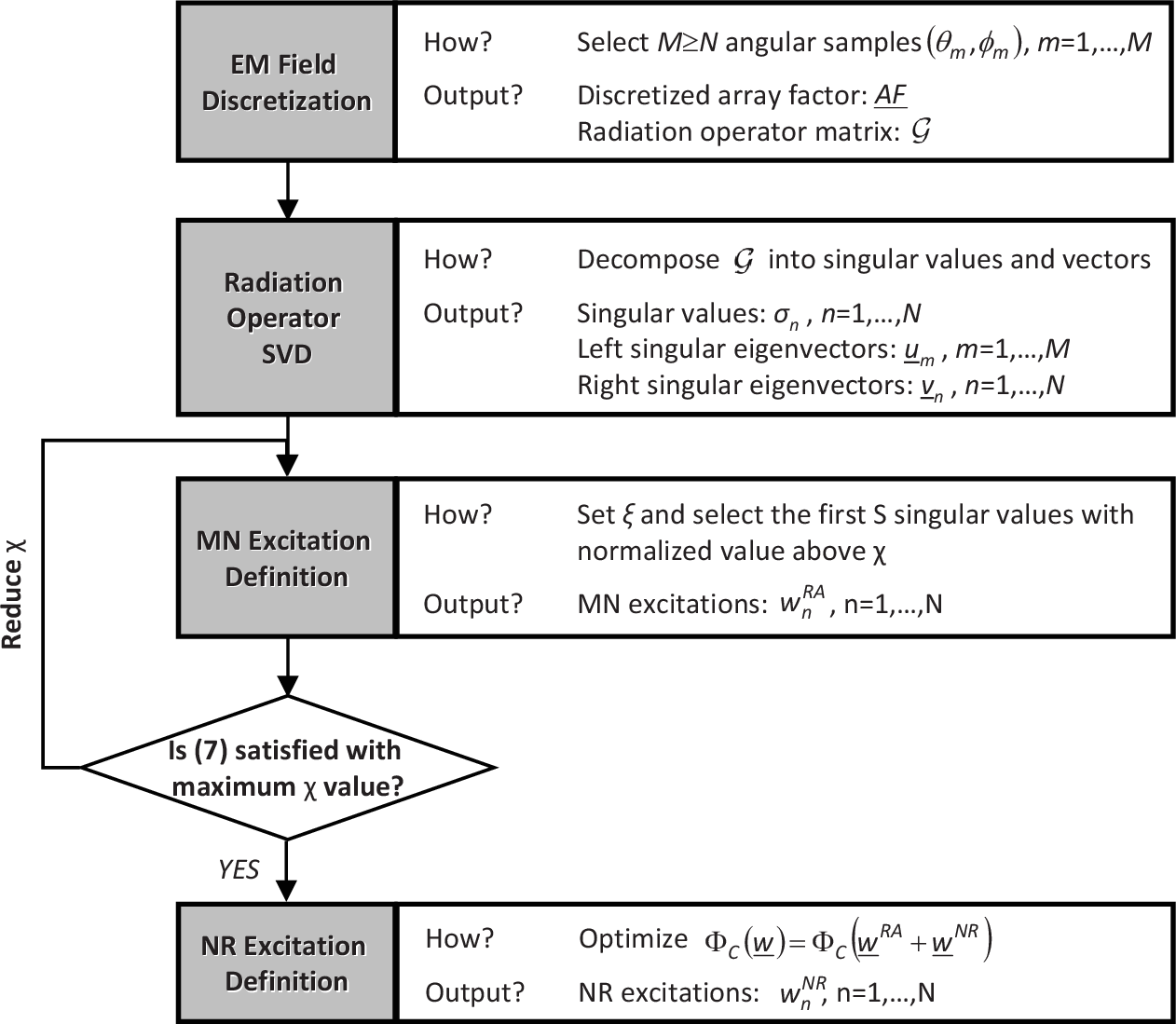}\tabularnewline
\end{tabular}\end{center}

\begin{center}~\vfill\end{center}

\begin{center}\textbf{Fig. 2 - Poli} \textbf{\emph{et al.,}} {}``Inverse
Source Method for Constrained Phased Array Synthesis ...''\end{center}

\newpage
\begin{center}~\vfill\end{center}

\begin{center}\begin{tabular}{cc}
\includegraphics[%
  width=0.50\columnwidth]{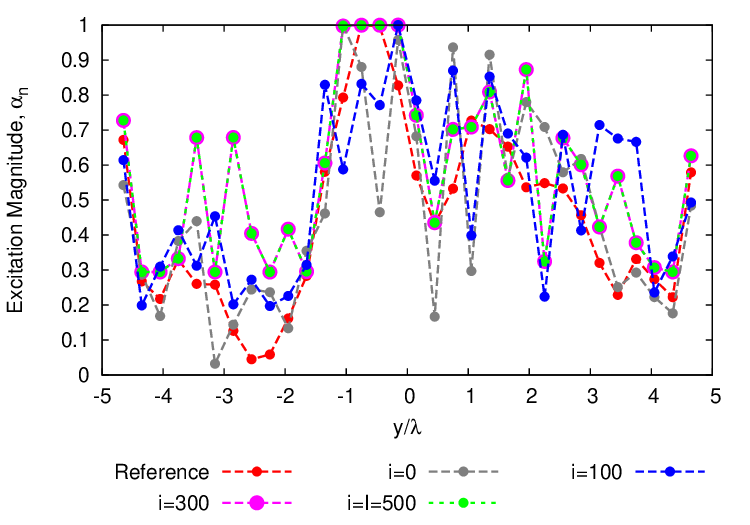}&
\includegraphics[%
  width=0.50\columnwidth]{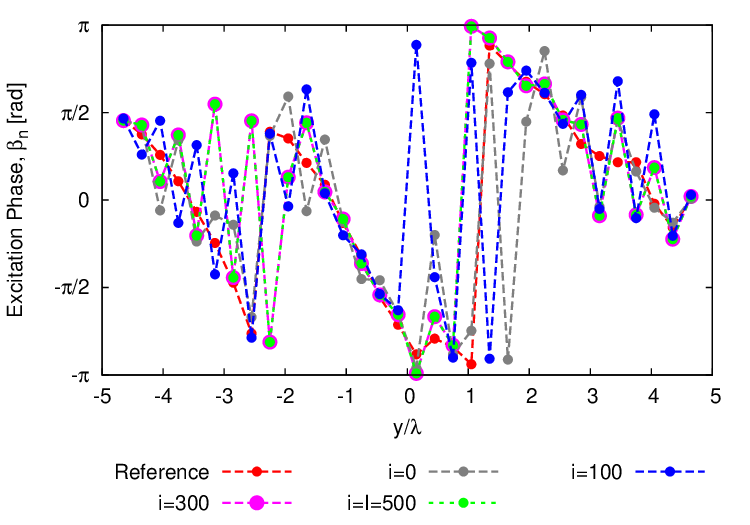}\tabularnewline
(\emph{a})&
(\emph{b})\tabularnewline
\includegraphics[%
  width=0.50\columnwidth]{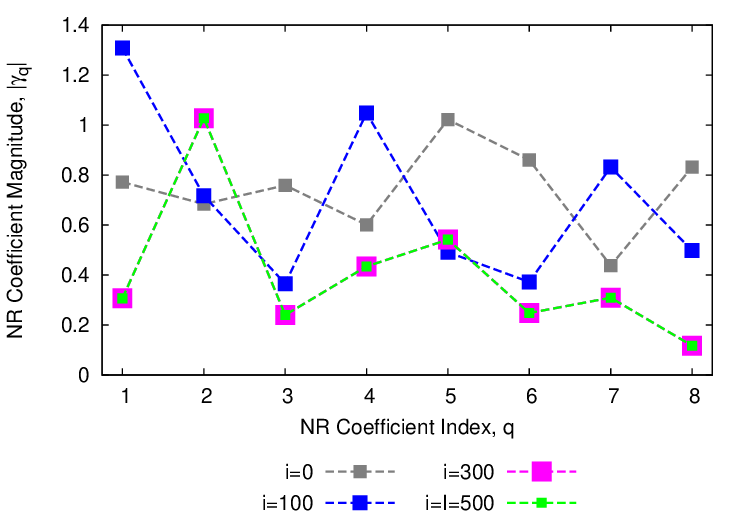}&
\includegraphics[%
  width=0.50\columnwidth]{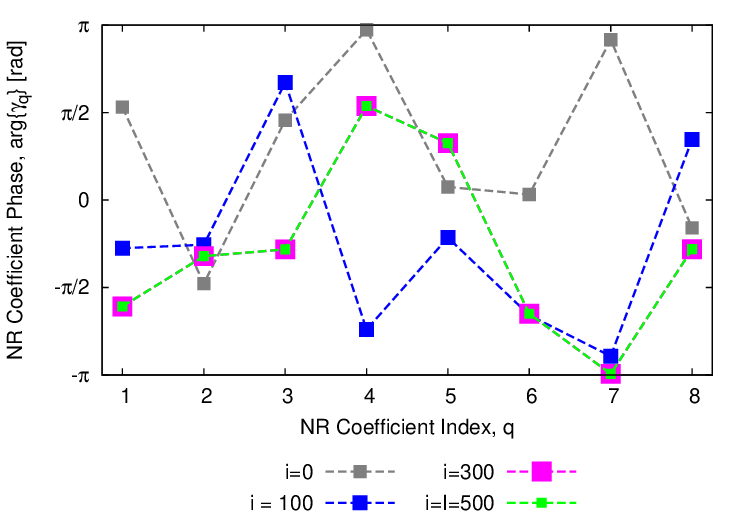}\tabularnewline
(\emph{c})&
(\emph{d})\tabularnewline
\multicolumn{2}{c}{\includegraphics[%
  width=0.50\columnwidth]{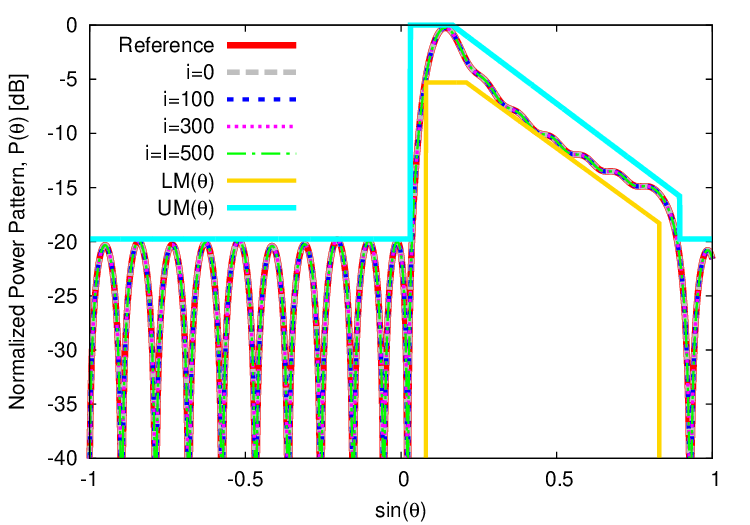}}\tabularnewline
\multicolumn{2}{c}{(\emph{e})}\tabularnewline
\end{tabular}\end{center}

\begin{center}~\vfill\end{center}

\begin{center}\textbf{Fig. 3 - Poli} \textbf{\emph{et al.,}} {}``Inverse
Source Method for Constrained Phased Array Synthesis ...''\end{center}

\newpage
\begin{center}~\vfill\end{center}

\begin{center}\begin{tabular}{cc}
\includegraphics[%
  width=0.50\columnwidth]{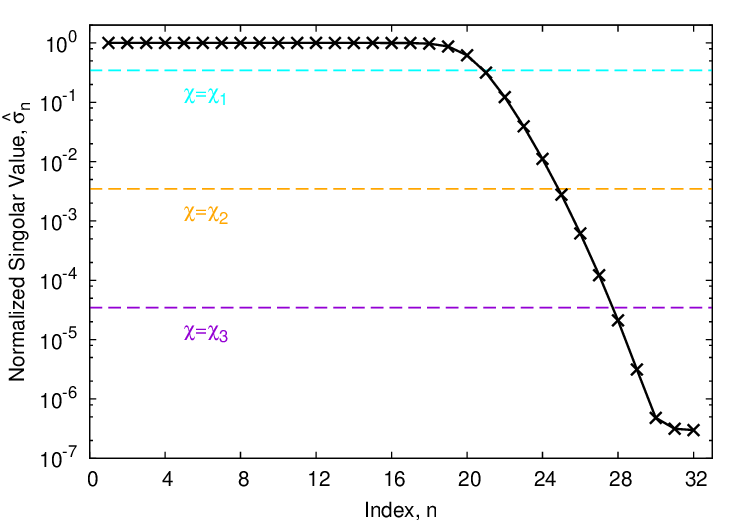}&
\includegraphics[%
  width=0.50\columnwidth]{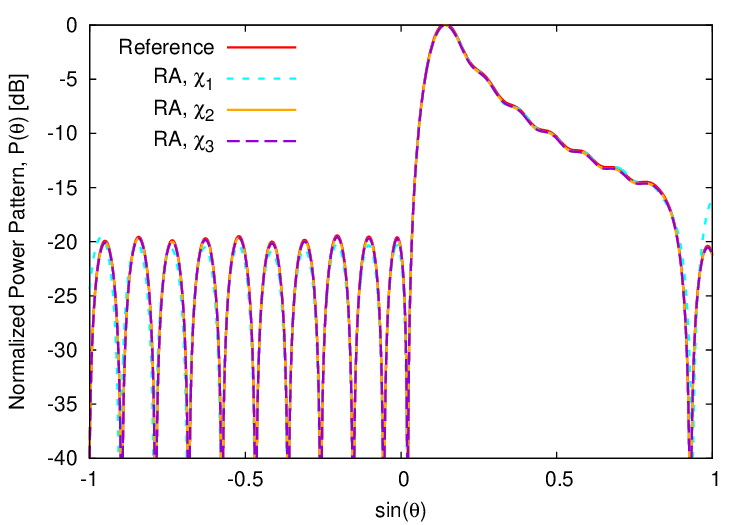}\tabularnewline
(\emph{a})&
(\emph{b})\tabularnewline
\includegraphics[%
  width=0.50\columnwidth]{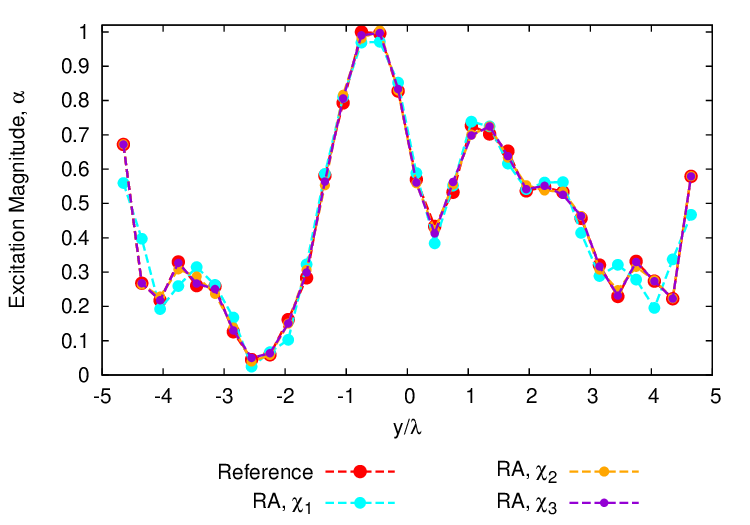}&
\includegraphics[%
  width=0.50\columnwidth]{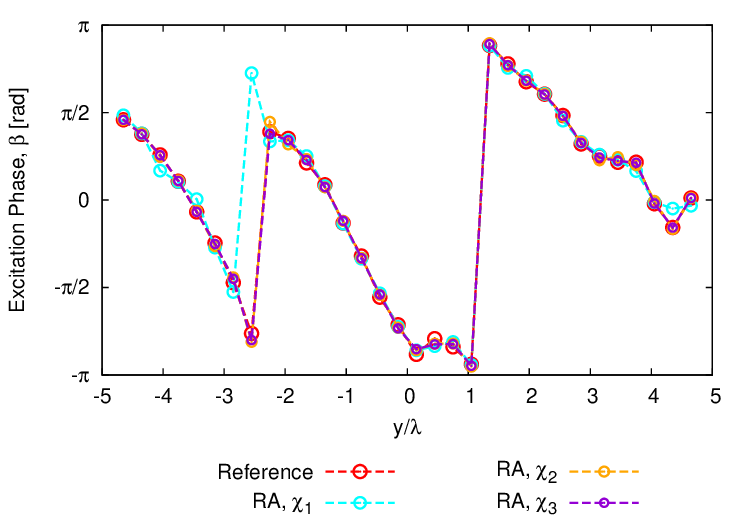}\tabularnewline
(\emph{c})&
(\emph{d})\tabularnewline
\end{tabular}\end{center}

\begin{center}~\vfill\end{center}

\begin{center}\textbf{Fig. 4 - Poli} \textbf{\emph{et al.,}} {}``Inverse
Source Method for Constrained Phased Array Synthesis ...''\end{center}

\newpage
\begin{center}~\vfill\end{center}

\begin{center}\begin{tabular}{c}
\includegraphics[%
  width=0.80\columnwidth]{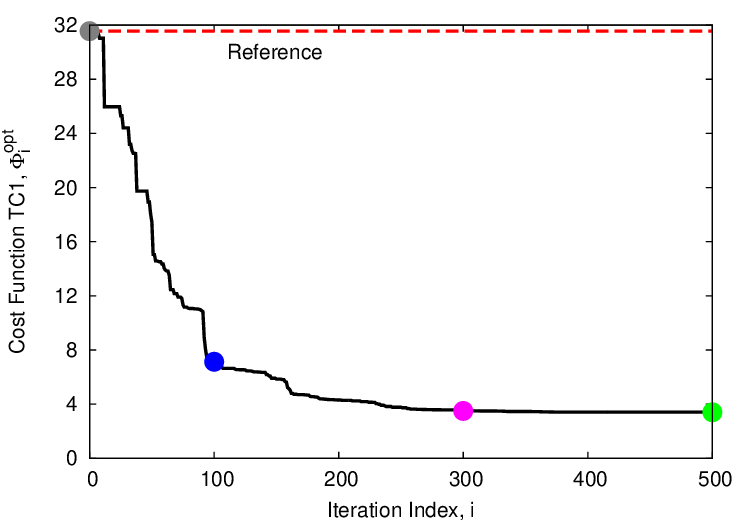}\tabularnewline
\end{tabular}\end{center}

\begin{center}~\vfill\end{center}

\begin{center}\textbf{Fig. 5 - Poli} \textbf{\emph{et al.,}} {}``Inverse
Source Method for Constrained Phased Array Synthesis ...''\end{center}

\newpage
\begin{center}~\vfill\end{center}

\begin{center}\begin{tabular}{cc}
\includegraphics[%
  width=0.50\columnwidth]{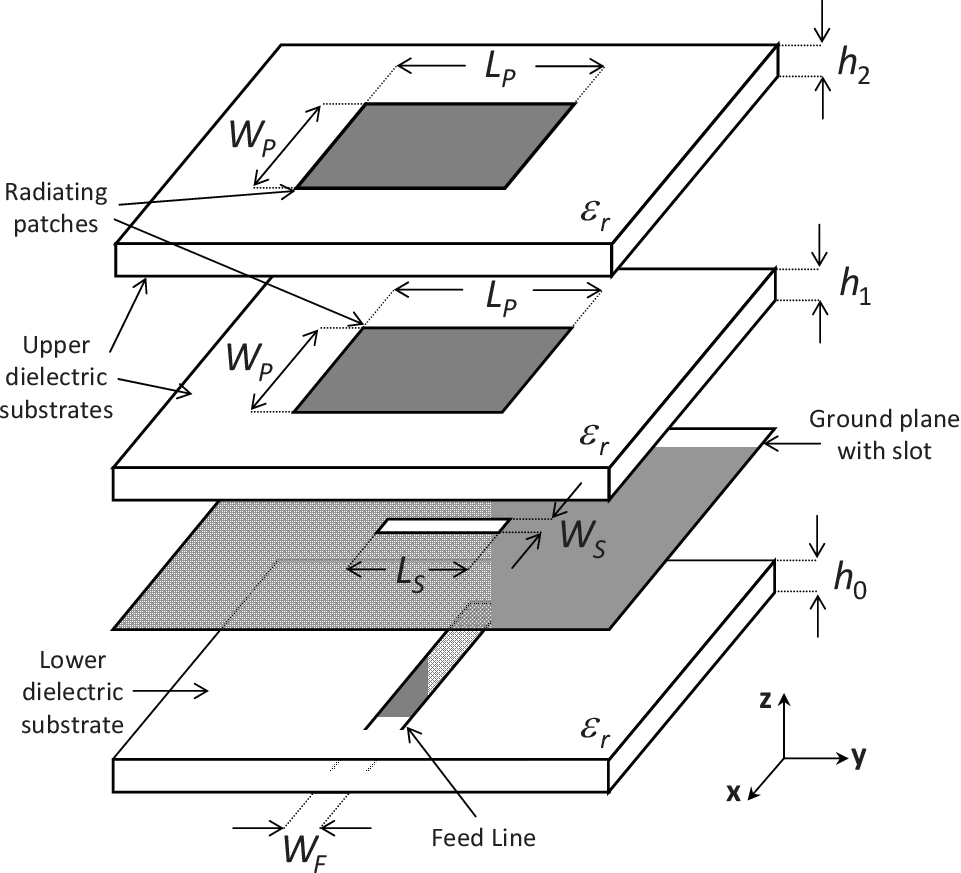}&
\includegraphics[%
  width=0.30\columnwidth]{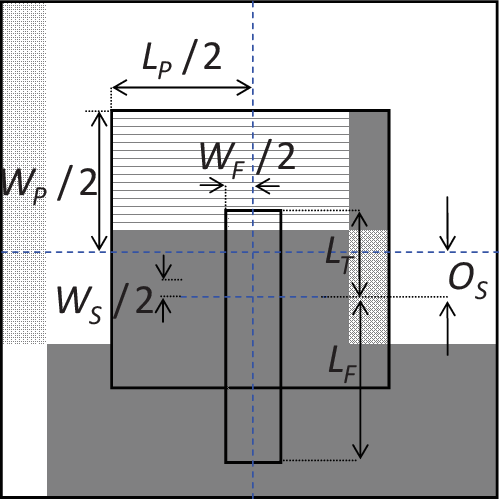}\tabularnewline
(\emph{a})&
(\emph{b})\tabularnewline
\multicolumn{2}{c}{\includegraphics[%
  width=1.0\columnwidth]{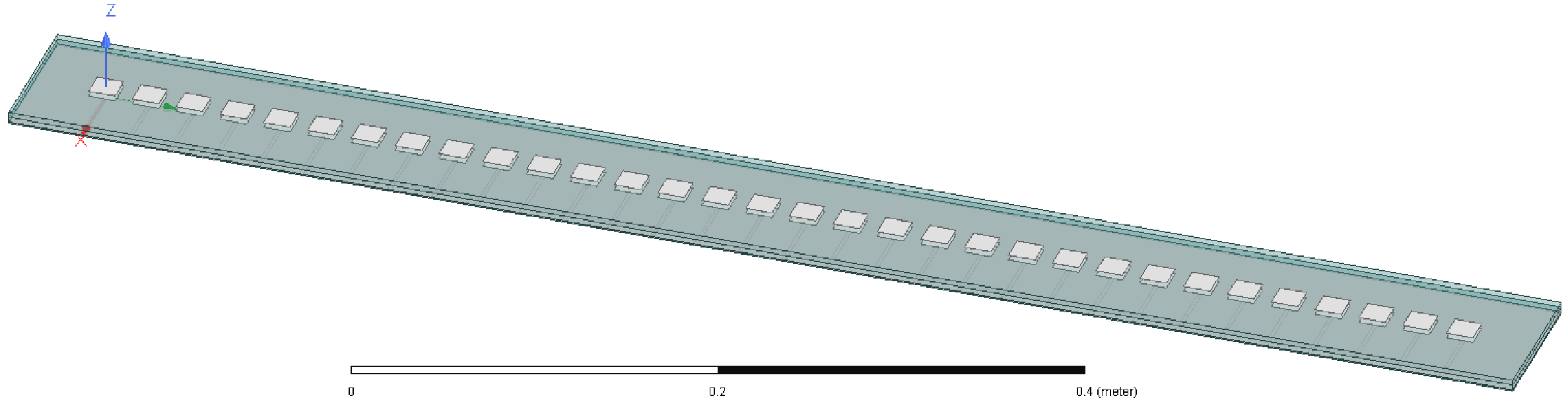}}\tabularnewline
\multicolumn{2}{c}{(\emph{c})}\tabularnewline
\end{tabular}\end{center}

\begin{center}~\vfill\end{center}

\begin{center}\textbf{Fig. 6 - Poli} \textbf{\emph{et al.,}} {}``Inverse
Source Method for Constrained Phased Array Synthesis ...''\end{center}

\newpage
\begin{center}~\vfill\end{center}

\begin{center}\begin{tabular}{c}
\includegraphics[%
  width=0.80\columnwidth]{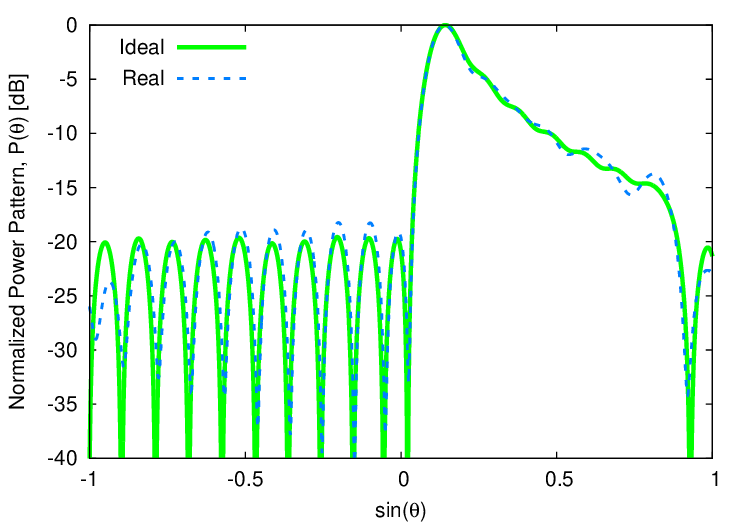}\tabularnewline
\end{tabular}\end{center}

\begin{center}~\vfill\end{center}

\begin{center}\textbf{Fig. 7 - Poli} \textbf{\emph{et al.,}} {}``Inverse
Source Method for Constrained Phased Array Synthesis ...''\end{center}

\newpage
\begin{center}~\vfill\end{center}

\begin{center}\begin{tabular}{c}
\includegraphics[%
  width=0.60\columnwidth]{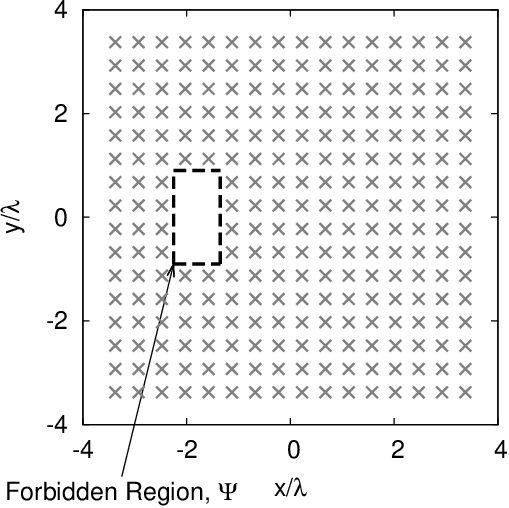}\tabularnewline
\end{tabular}\end{center}

\begin{center}~\vfill\end{center}

\begin{center}\textbf{Fig. 8 - Poli} \textbf{\emph{et al.,}} {}``Inverse
Source Method for Constrained Phased Array Synthesis ...''\end{center}

\newpage
\begin{center}~\vfill\end{center}

\begin{center}\begin{tabular}{cc}
\includegraphics[%
  width=0.45\columnwidth]{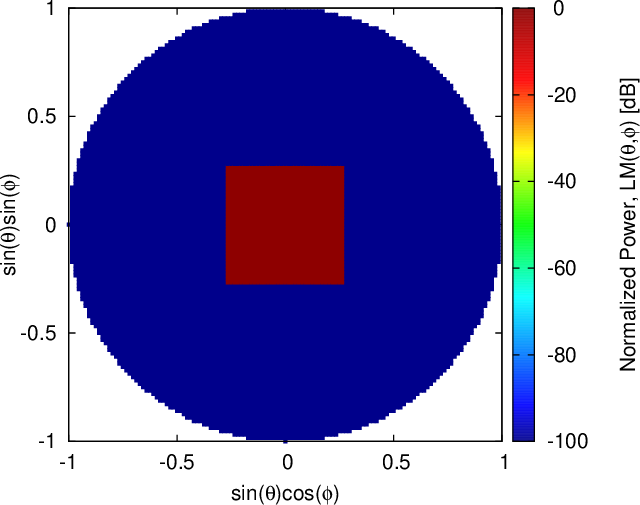}&
\includegraphics[%
  width=0.45\columnwidth]{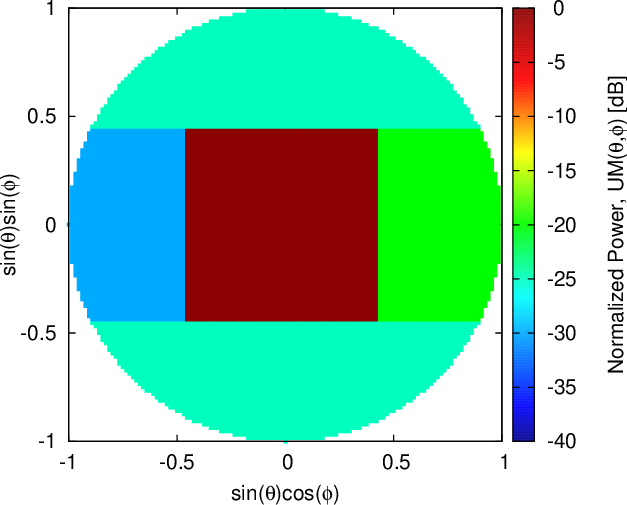}\tabularnewline
(\emph{a})&
(\emph{b})\tabularnewline
\includegraphics[%
  width=0.45\columnwidth]{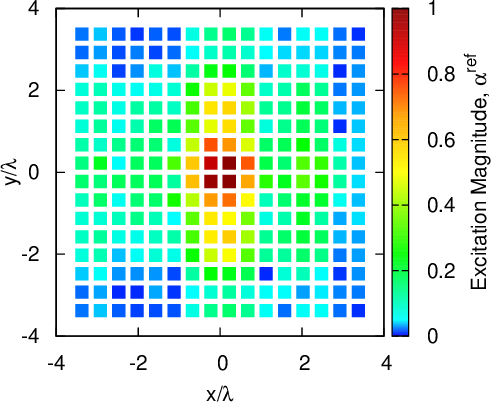}&
\includegraphics[%
  width=0.45\columnwidth]{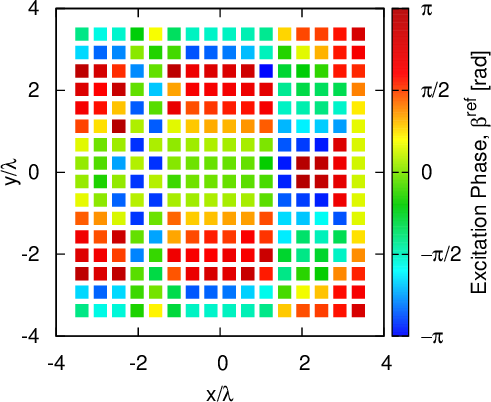}\tabularnewline
(\emph{c})&
(\emph{d})\tabularnewline
\multicolumn{2}{c}{\includegraphics[%
  width=0.45\columnwidth]{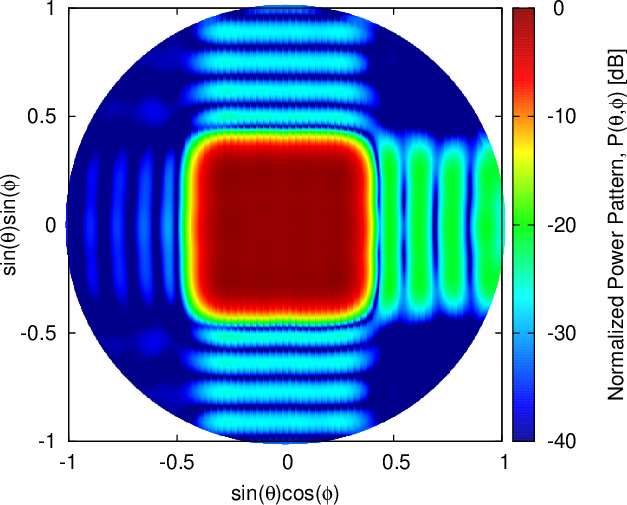}}\tabularnewline
\multicolumn{2}{c}{(\emph{e})}\tabularnewline
\end{tabular}\end{center}

\begin{center}~\vfill\end{center}

\begin{center}\textbf{Fig. 9 - Poli} \textbf{\emph{et al.,}} {}``Inverse
Source Method for Constrained Phased Array Synthesis ...''\end{center}

\newpage
\begin{center}~\vfill\end{center}

\begin{center}\begin{tabular}{c}
\includegraphics[%
  width=0.80\columnwidth]{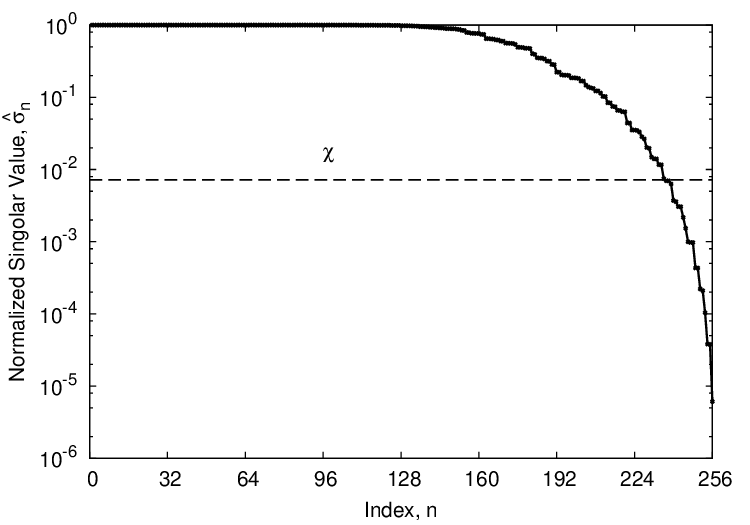}\tabularnewline
\end{tabular}\end{center}

\begin{center}~\vfill\end{center}

\begin{center}\textbf{Fig. 10 - Poli} \textbf{\emph{et al.,}} {}``Inverse
Source Method for Constrained Phased Array Synthesis ...''\end{center}

\newpage
\begin{center}~\vfill\end{center}

\begin{center}\begin{tabular}{c}
\includegraphics[%
  width=0.80\columnwidth]{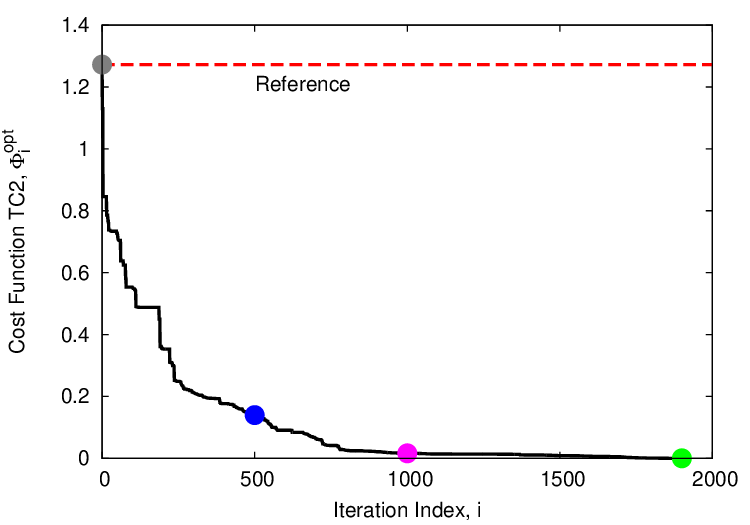}\tabularnewline
\end{tabular}\end{center}

\begin{center}~\vfill\end{center}

\begin{center}\textbf{Fig. 11 - Poli} \textbf{\emph{et al.,}} {}``Inverse
Source Method for Constrained Phased Array Synthesis ...''\end{center}

\newpage
\begin{center}~\vfill\end{center}

\begin{center}\begin{tabular}{cc}
\includegraphics[%
  width=0.50\columnwidth]{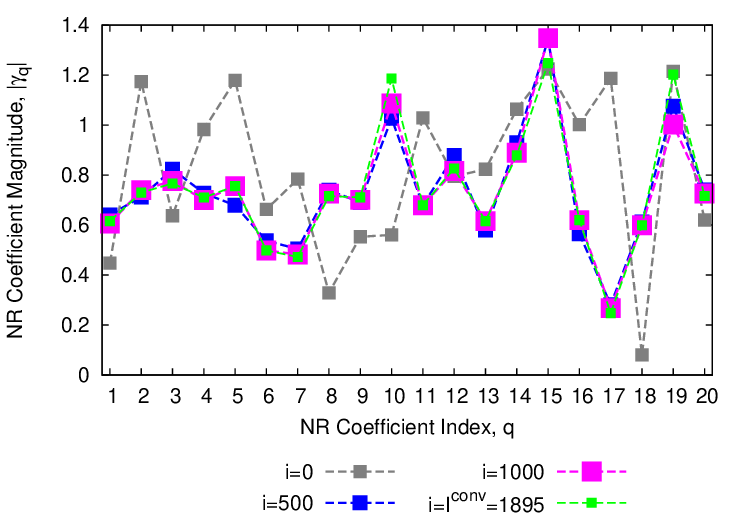}&
\includegraphics[%
  width=0.50\columnwidth]{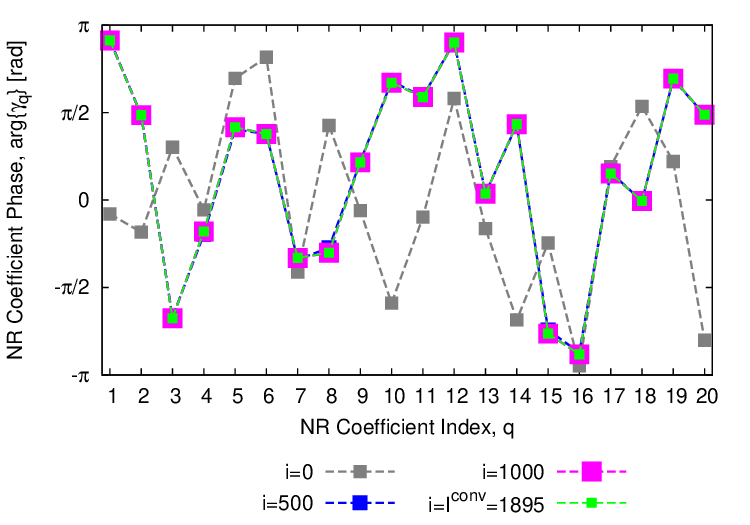}\tabularnewline
(\emph{a})&
(\emph{b})\tabularnewline
\includegraphics[%
  width=0.50\columnwidth]{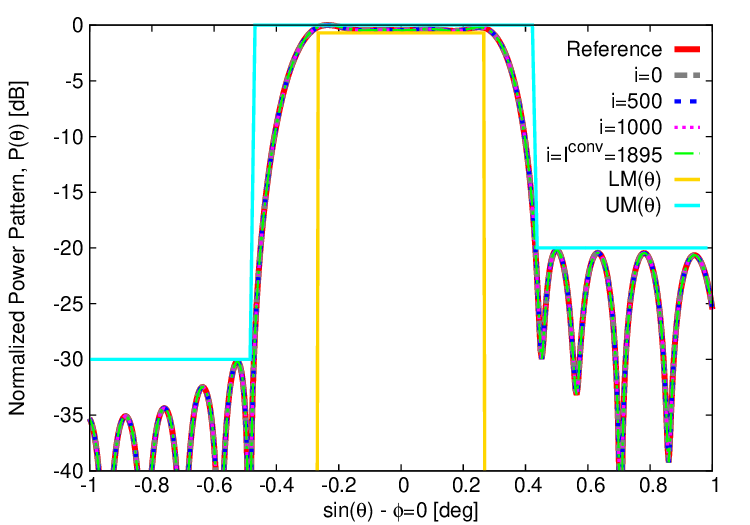}&
\includegraphics[%
  width=0.50\columnwidth]{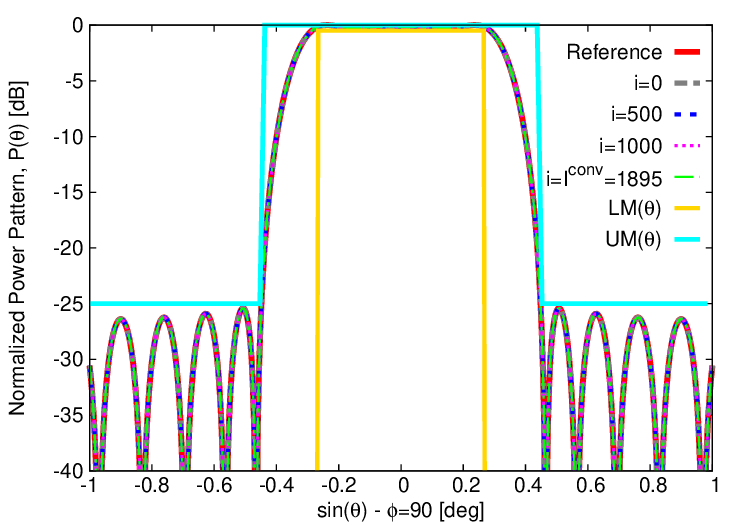}\tabularnewline
(\emph{c})&
(\emph{d})\tabularnewline
\end{tabular}\end{center}

\begin{center}~\vfill\end{center}

\begin{center}\textbf{Fig. 12 - Poli} \textbf{\emph{et al.,}} {}``Inverse
Source Method for Constrained Phased Array Synthesis ...''\end{center}
\newpage

\begin{center}\begin{sideways}
\begin{tabular}{cccc}
$i=0$&
$i=500$&
$i=1000$&
$i=I^{conv}=1895$\tabularnewline
\includegraphics[%
  width=0.33\columnwidth]{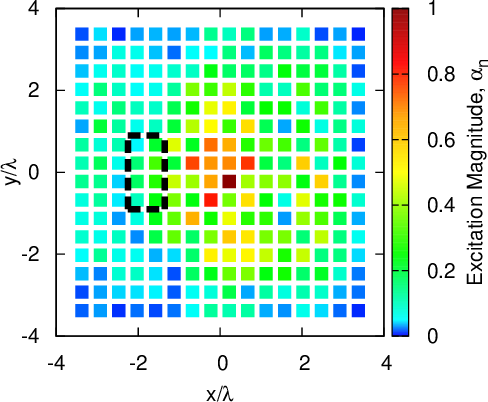}&
\includegraphics[%
  width=0.33\columnwidth]{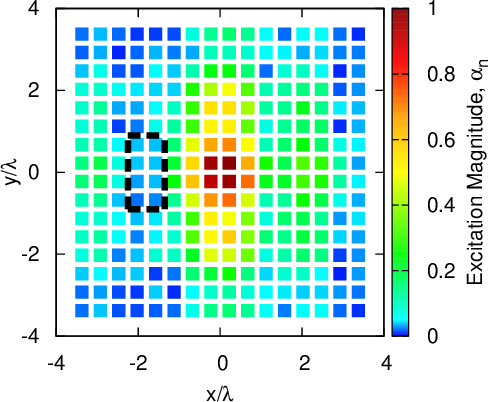}&
\includegraphics[%
  width=0.33\columnwidth]{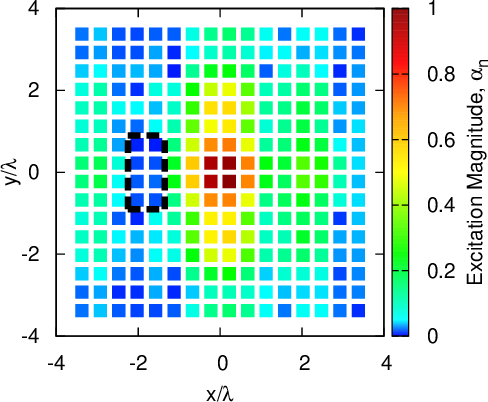}&
\includegraphics[%
  width=0.33\columnwidth]{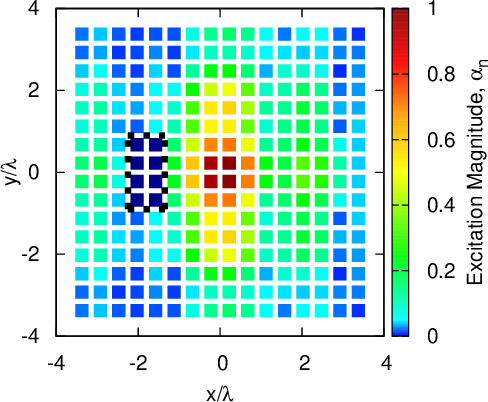}\tabularnewline
(\emph{a})&
(\emph{b})&
(\emph{c})&
(\emph{d})\tabularnewline
\includegraphics[%
  width=0.33\columnwidth]{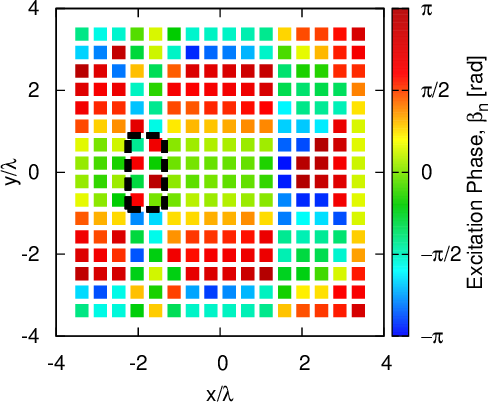}&
\includegraphics[%
  width=0.33\columnwidth]{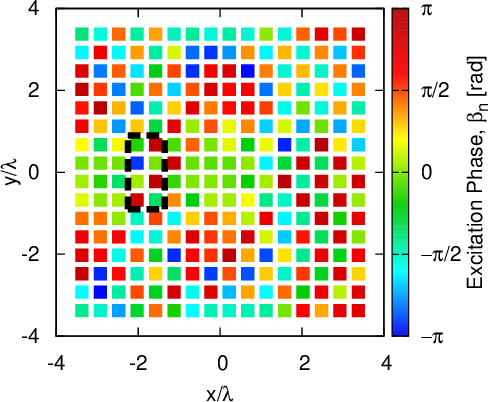}&
\includegraphics[%
  width=0.33\columnwidth]{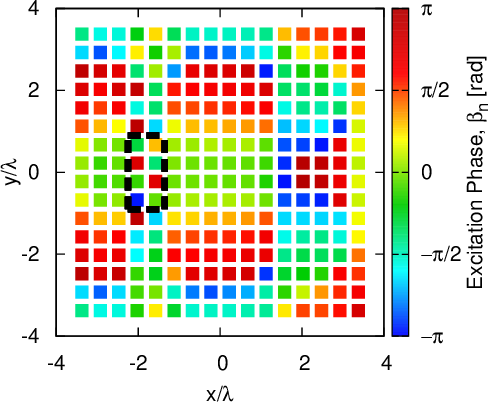}&
\includegraphics[%
  width=0.33\columnwidth]{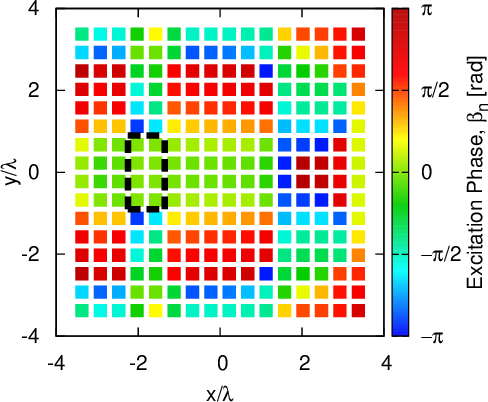}\tabularnewline
(\emph{e})&
(\emph{f})&
(\emph{g})&
(\emph{h})\tabularnewline
\end{tabular}
\end{sideways}\end{center}

\begin{center}\textbf{Fig. 13 - Poli} \textbf{\emph{et al.,}} {}``Inverse
Source Method for Constrained Phased Array Synthesis ...''\end{center}

\newpage
\begin{center}~\vfill\end{center}

\begin{center}\begin{tabular}{c}
\includegraphics[%
  width=0.80\columnwidth]{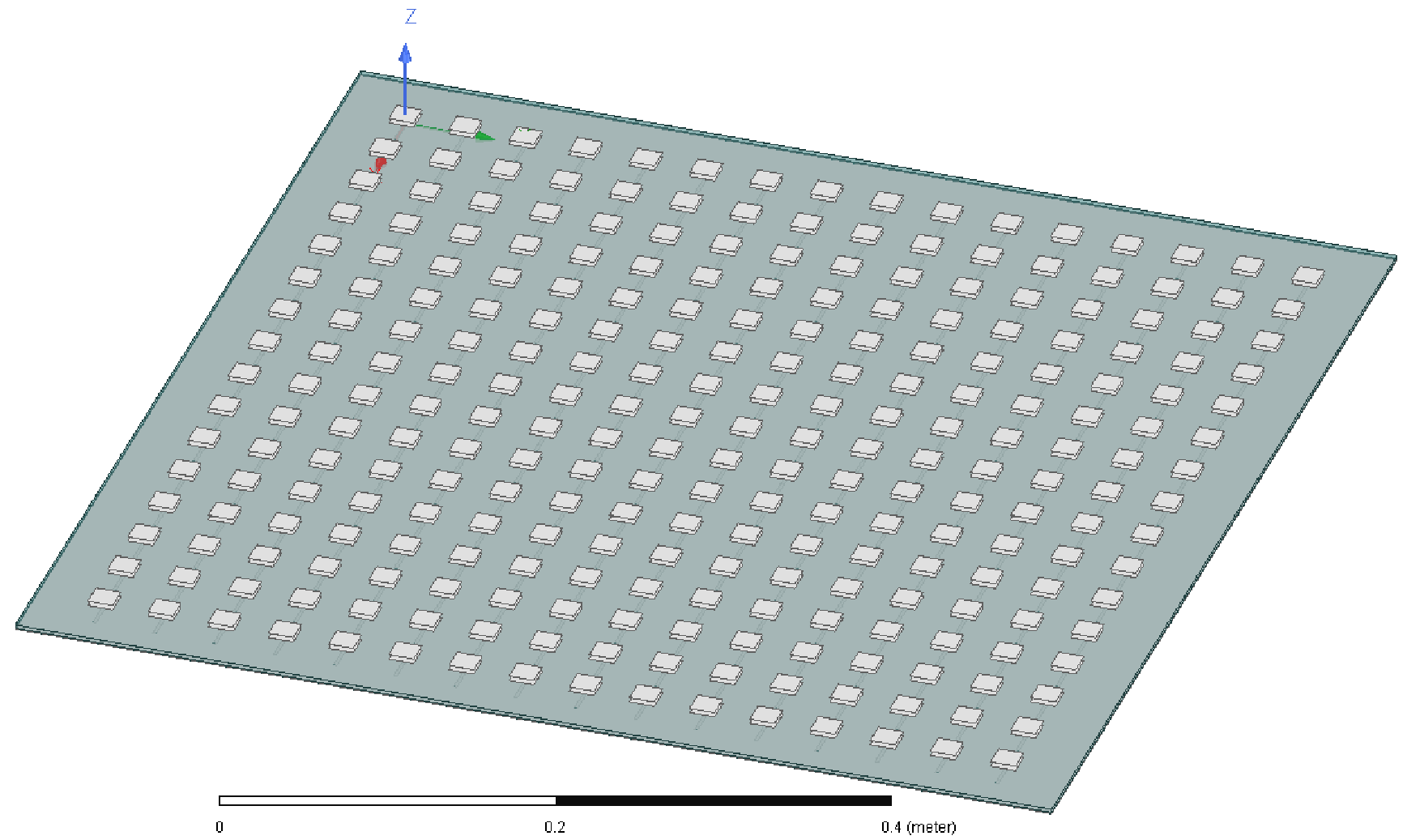}\tabularnewline
\end{tabular}\end{center}

\begin{center}~\vfill\end{center}

\begin{center}\textbf{Fig. 14 - Poli} \textbf{\emph{et al.,}} {}``Inverse
Source Method for Constrained Phased Array Synthesis ...''\end{center}

\newpage
\begin{center}~\vfill\end{center}

\begin{center}\begin{tabular}{c}
\includegraphics[%
  width=0.70\columnwidth]{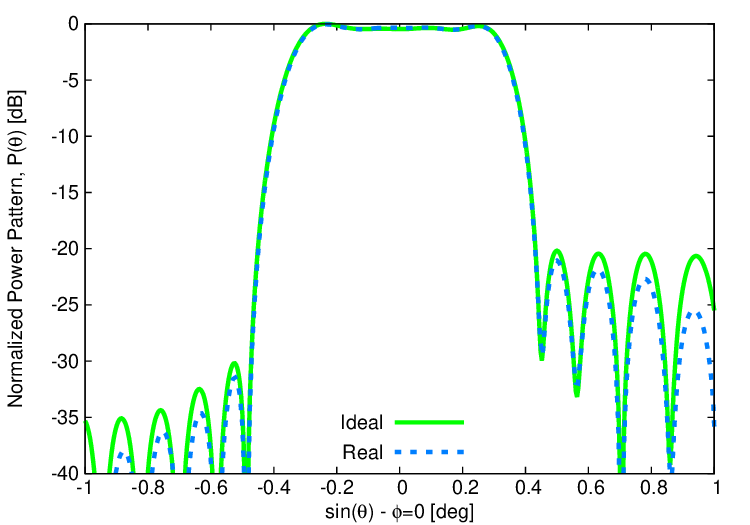}\tabularnewline
(\emph{a})\tabularnewline
\includegraphics[%
  width=0.70\columnwidth]{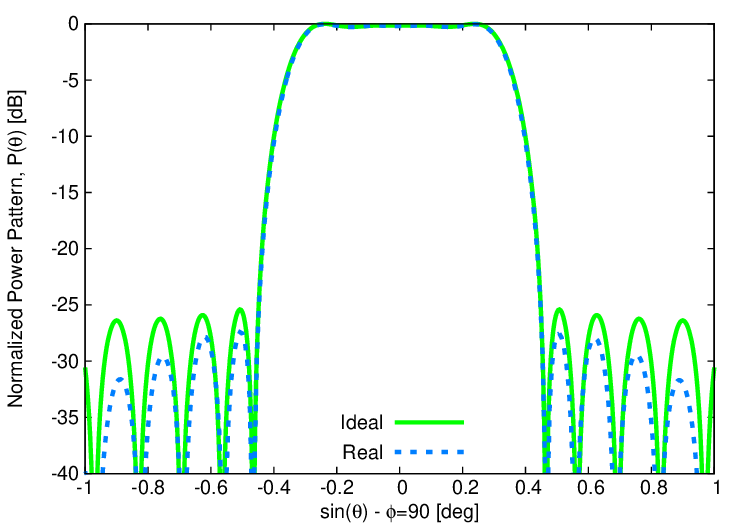}\tabularnewline
(\emph{b})\tabularnewline
\end{tabular}\end{center}

\begin{center}~\vfill\end{center}

\begin{center}\textbf{Fig. 15 - Poli} \textbf{\emph{et al.,}} {}``Inverse
Source Method for Constrained Phased Array Synthesis ...''\end{center}

\newpage
\begin{center}~\vfill\end{center}

\begin{center}\begin{tabular}{cc}
\includegraphics[%
  width=0.45\columnwidth]{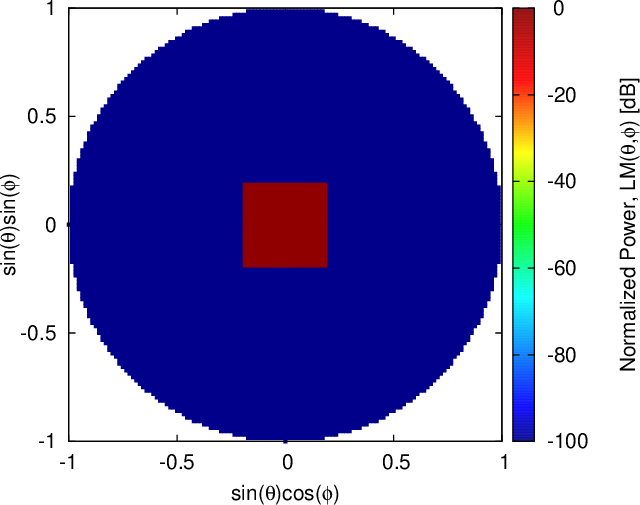}&
\includegraphics[%
  width=0.45\columnwidth]{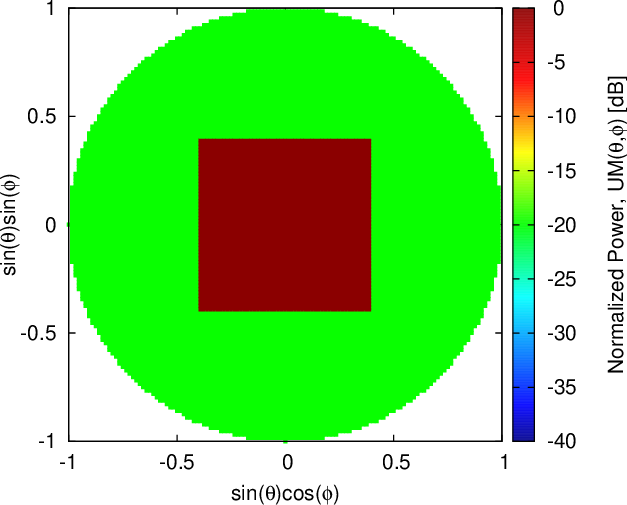}\tabularnewline
(\emph{a})&
(\emph{b})\tabularnewline
\includegraphics[%
  width=0.45\columnwidth]{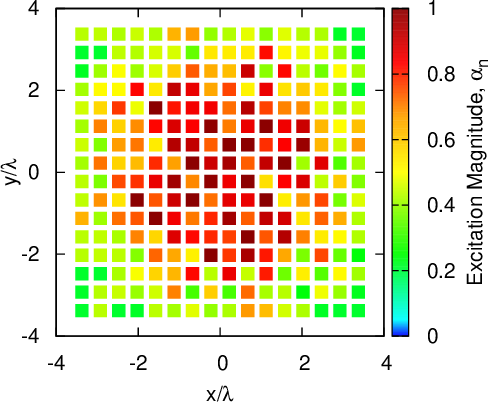}&
\includegraphics[%
  width=0.45\columnwidth]{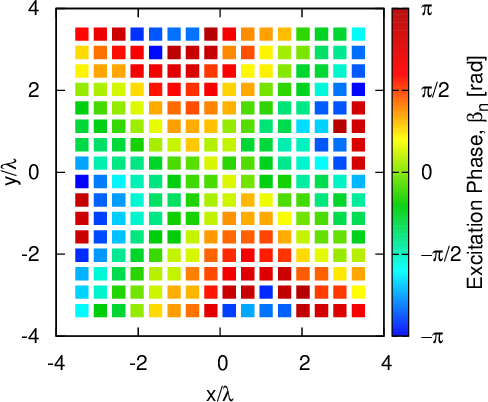}\tabularnewline
(\emph{c})&
(\emph{d})\tabularnewline
\multicolumn{2}{c}{\includegraphics[%
  width=0.45\columnwidth]{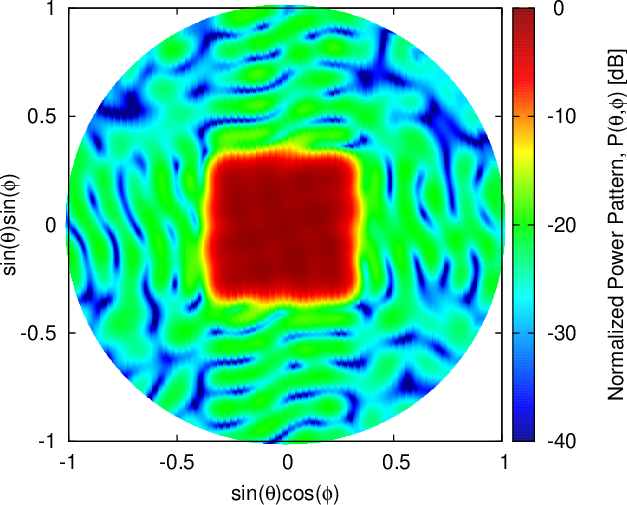}}\tabularnewline
\multicolumn{2}{c}{(\emph{e})}\tabularnewline
\end{tabular}\end{center}

\begin{center}~\vfill\end{center}

\begin{center}\textbf{Fig. 16 - Poli} \textbf{\emph{et al.,}} {}``Inverse
Source Method for Constrained Phased Array Synthesis ...''\end{center}

\newpage
\begin{center}~\vfill\end{center}

\begin{center}\begin{tabular}{c}
\includegraphics[%
  width=0.80\columnwidth]{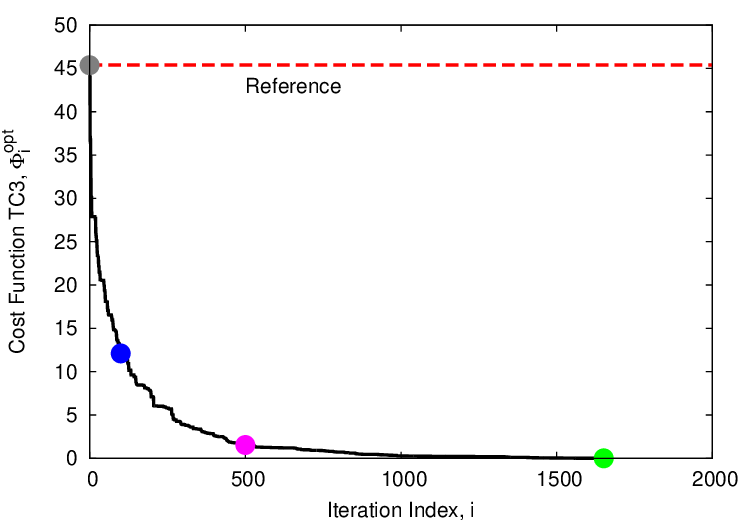}\tabularnewline
\end{tabular}\end{center}

\begin{center}~\vfill\end{center}

\begin{center}\textbf{Fig. 17 - Poli} \textbf{\emph{et al.,}} {}``Inverse
Source Method for Constrained Phased Array Synthesis ...''\end{center}

\newpage
\begin{center}~\vfill\end{center}

\begin{center}\begin{tabular}{cc}
\includegraphics[%
  width=0.50\columnwidth]{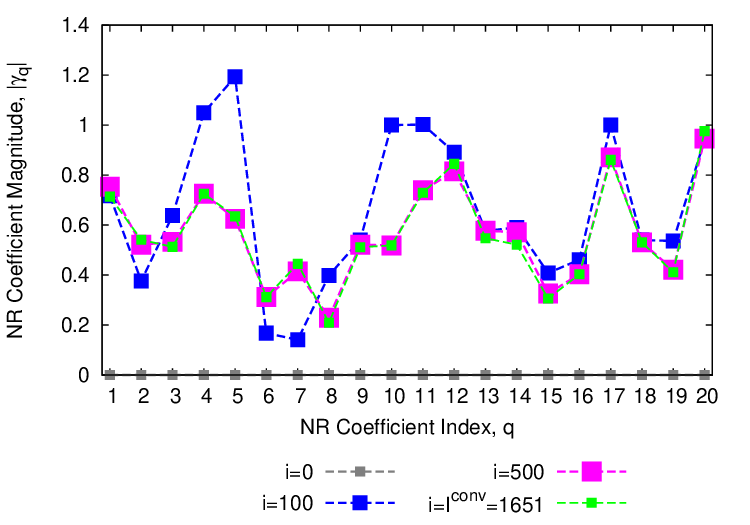}&
\includegraphics[%
  width=0.50\columnwidth]{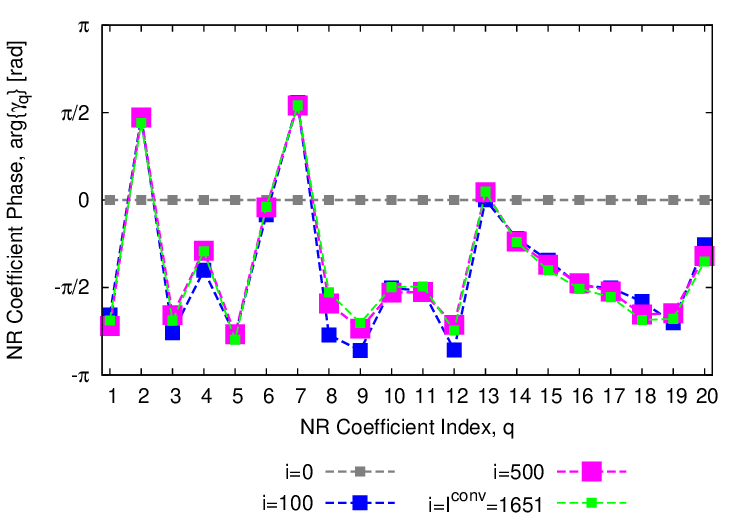}\tabularnewline
(\emph{a})&
(\emph{b})\tabularnewline
\includegraphics[%
  width=0.50\columnwidth]{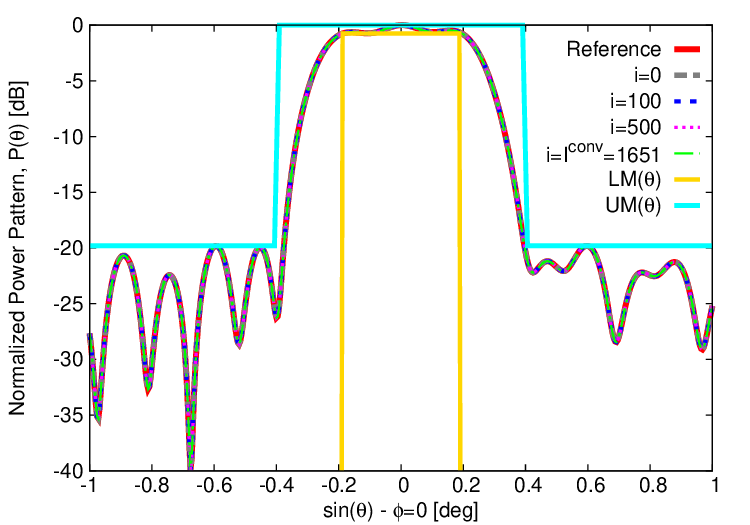}&
\includegraphics[%
  width=0.50\columnwidth]{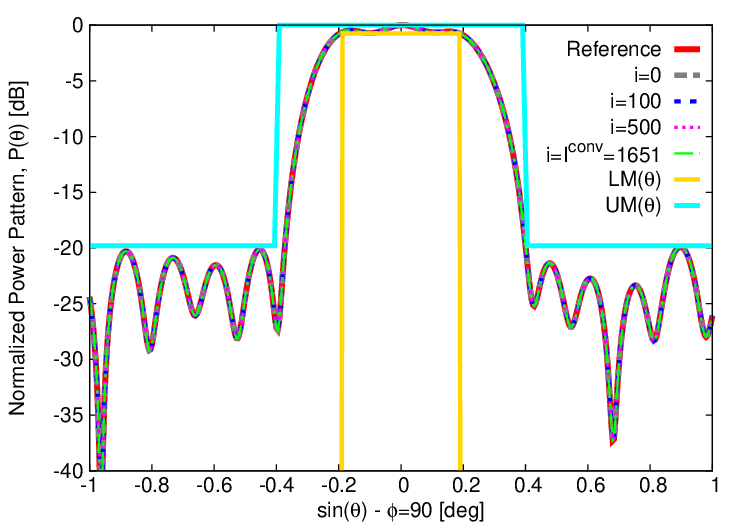}\tabularnewline
(\emph{c})&
(\emph{d})\tabularnewline
\end{tabular}\end{center}

\begin{center}~\vfill\end{center}

\begin{center}\textbf{Fig. 18 - Poli} \textbf{\emph{et al.,}} {}``Inverse
Source Method for Constrained Phased Array Synthesis ...''\end{center}

\newpage
\begin{center}~\vfill\end{center}

\begin{center}\begin{tabular}{ccc}
\begin{sideways}
~~~~~~$i=0$%
\end{sideways}&
\includegraphics[%
  width=0.32\columnwidth]{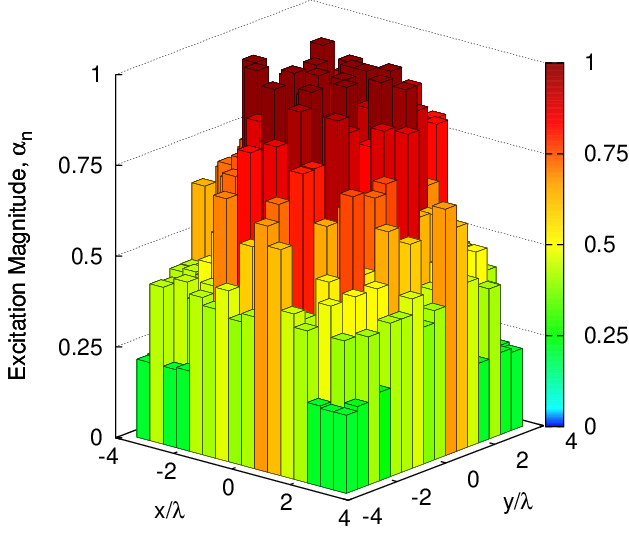}&
\includegraphics[%
  width=0.32\columnwidth]{Fig.16d.eps}\tabularnewline
&
(\emph{a})&
(\emph{e})\tabularnewline
\begin{sideways}
~~~~~~$i=100$%
\end{sideways}&
\includegraphics[%
  width=0.32\columnwidth]{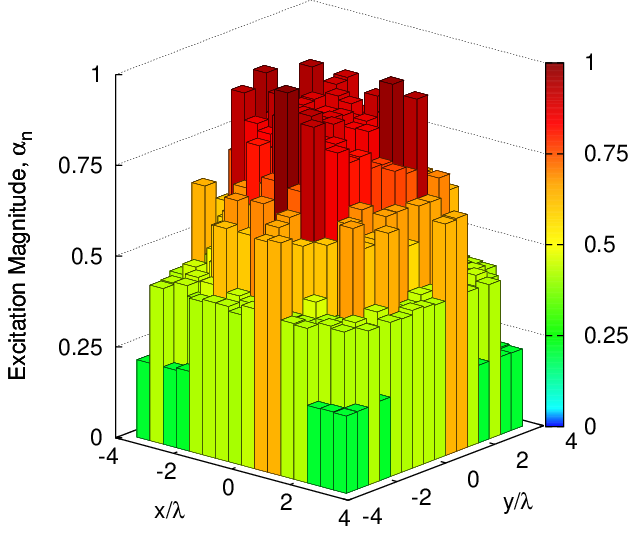}&
\includegraphics[%
  width=0.32\columnwidth]{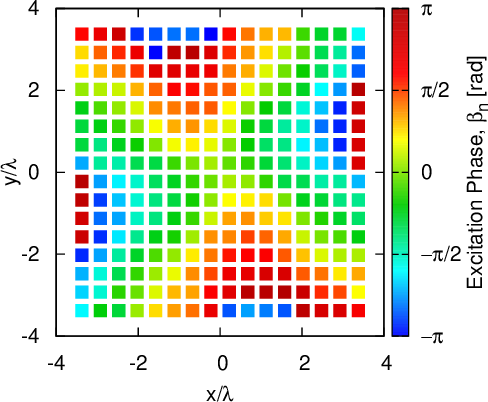}\tabularnewline
&
(\emph{b})&
(\emph{f})\tabularnewline
\begin{sideways}
~~~~~~$i=500$%
\end{sideways}&
\includegraphics[%
  width=0.32\columnwidth]{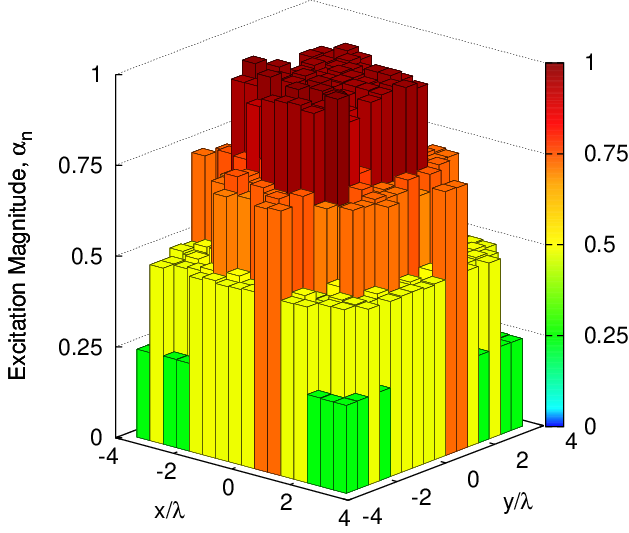}&
\includegraphics[%
  width=0.32\columnwidth]{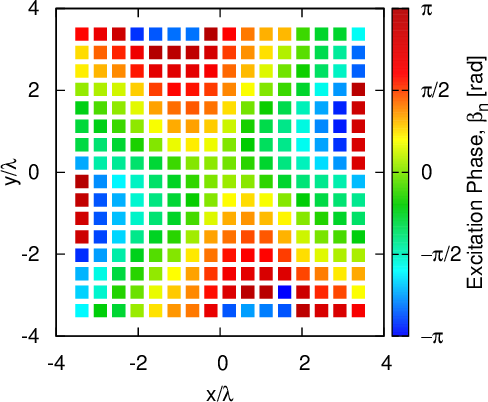}\tabularnewline
&
(\emph{c})&
(\emph{g})\tabularnewline
\begin{sideways}
~~~~~~$i=I^{conv}=1651$%
\end{sideways}&
\includegraphics[%
  width=0.32\columnwidth]{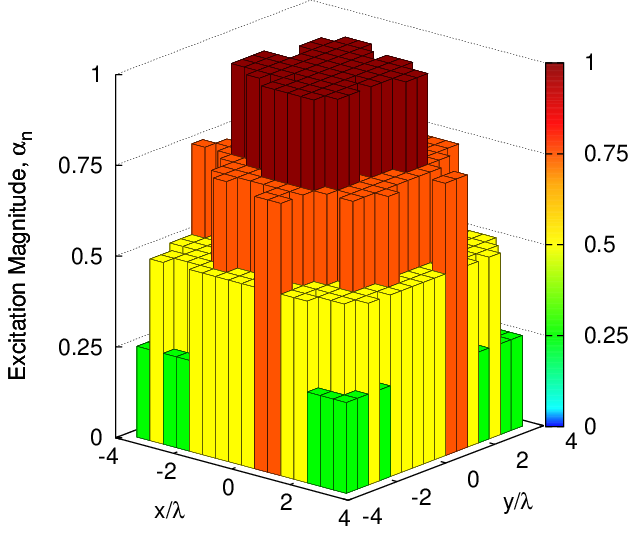}&
\includegraphics[%
  width=0.32\columnwidth]{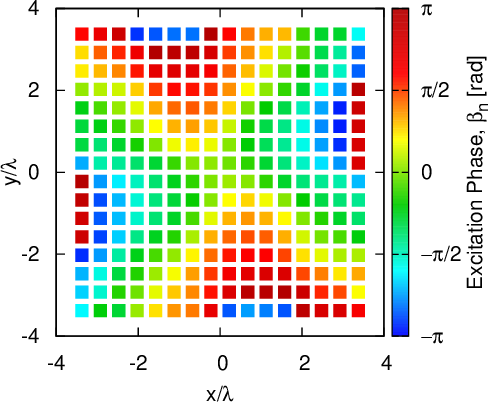}\tabularnewline
&
(\emph{d})&
(\emph{h})\tabularnewline
\end{tabular}\end{center}

\begin{center}~\vfill\end{center}

\begin{center}\textbf{Fig. 19 - Poli} \textbf{\emph{et al.,}} {}``Inverse
Source Method for Constrained Phased Array Synthesis ...''\end{center}

\newpage
\begin{center}~\vfill\end{center}

\begin{center}\begin{tabular}{c}
\includegraphics[%
  width=0.70\columnwidth]{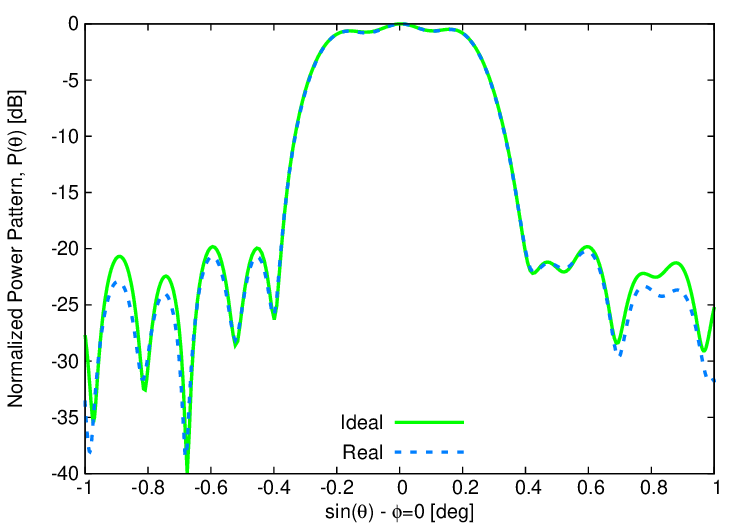}\tabularnewline
(\emph{a})\tabularnewline
\includegraphics[%
  width=0.70\columnwidth]{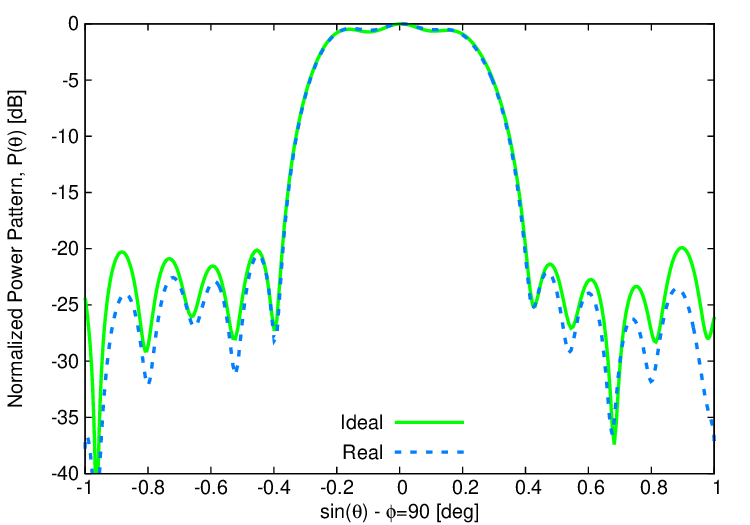}\tabularnewline
(\emph{b})\tabularnewline
\end{tabular}\end{center}

\begin{center}~\vfill\end{center}

\begin{center}\textbf{Fig. 20 - Poli} \textbf{\emph{et al.,}} {}``Inverse
Source Method for Constrained Phased Array Synthesis ...''\end{center}

\newpage
\begin{center}~\vfill\end{center}

\begin{center}\begin{tabular}{|c|c|}
\hline 
\emph{Parameter}&
\emph{Value}\tabularnewline
\hline
\hline 
$h_{0}$&
$7.60\times10^{-4}$ {[}m{]}\tabularnewline
\hline 
$h_{1}$&
$4.56\times10^{-3}$ {[}m{]}\tabularnewline
\hline 
$h_{2}$&
$4.56\times10^{-3}$ {[}m{]}\tabularnewline
\hline 
$h_{p}=h_{G}$&
$3.50\times10^{-5}$ {[}m{]}\tabularnewline
\hline 
$LP=WP$&
$2.04\times10^{-2}$ {[}m{]}\tabularnewline
\hline 
$WF$&
$4.08\times10^{-3}$ {[}m{]}\tabularnewline
\hline 
$L_{F}$&
$4.66\times10^{-3}$ {[}m{]}\tabularnewline
\hline 
$L_{T}$&
$3.51\times10^{-3}$ {[}m{]}\tabularnewline
\hline 
$O_{S}$&
$1.60\times10^{-3}$ {[}m{]}\tabularnewline
\hline 
$WS$&
$5.94\times10^{-3}$ {[}m{]}\tabularnewline
\hline 
$LS$&
$2.04\times10^{-2}$ {[}m{]}\tabularnewline
\hline 
$\varepsilon_{r}$&
$3.0$\tabularnewline
\hline
$\tan\delta$&
$1.6\times10^{-3}$\tabularnewline
\hline
\end{tabular}\end{center}

\begin{center}~\vfill\end{center}

\begin{center}\textbf{Tab. I - Poli} \textbf{\emph{et al.,}} {}``Inverse
Source Method for Constrained Phased Array Synthesis ...''\end{center}
\end{document}